\documentclass{emulateapj} 
 
\newcommand{\lsim}{\raisebox{-0.13cm}{~\shortstack{$<$ \\[-0.07cm] $\sim$}}~}
\newcommand{\gsim}{\raisebox{-0.13cm}{~\shortstack{$>$ \\[-0.07cm] $\sim$}}~}
\newcommand{\tfm}{$\rm 24 \, \mu m \,$}
\newcommand{\nLntfm}{$\nu L_\nu^{\rm 24 \, \mu m} \,$}

\newcommand{\oii}{$\rm [OII] \, \lambda 3727 \,$}
\newcommand{\hb}{$\rm H\beta \, \lambda 4861 \,$}
\newcommand{\oiii}{$\rm [OIII] \, \lambda 5007 \,$}
\newcommand{\ha}{$\rm H\alpha \, \lambda 6563 \,$}
\newcommand{\nii}{$\rm [NII] \, \lambda 6584 \,$}

\shorttitle{The optical spectra of \tfm galaxies in COSMOS: II}
\shortauthors{K. I. Caputi et al.}

\begin{document} 
 
 
\title{The optical spectra of {\em Spitzer} \tfm galaxies in the COSMOS field:  \\
II. Faint infrared sources in the \lowercase{z}COSMOS-bright 10\lowercase{k} catalogue}

 
\author{K. I. \ Caputi\altaffilmark{1,2}, 
S. J.  Lilly\altaffilmark{1}, 
H.  Aussel\altaffilmark{3},
E. Le Floc'h\altaffilmark{3},
D. Sanders\altaffilmark{4},
C. Maier\altaffilmark{1},
D. Frayer\altaffilmark{5},
C. M. Carollo\altaffilmark{1},
T. Contini\altaffilmark{6}, 
J.-P. Kneib\altaffilmark{7},
O. Le F\`evre\altaffilmark{7},
V. Mainieri\altaffilmark{8},
A. Renzini\altaffilmark{9},
M. Scodeggio\altaffilmark{10},
N. Scoville\altaffilmark{11},
G. Zamorani\altaffilmark{12}, 
S. Bardelli\altaffilmark{12},
M. Bolzonella\altaffilmark{12},
A. Bongiorno\altaffilmark{13},
G. Coppa\altaffilmark{12},
O. Cucciati\altaffilmark{7},
S. de la Torre\altaffilmark{7},
L. de Ravel\altaffilmark{7},
P. Franzetti\altaffilmark{10},
B. Garilli\altaffilmark{10},
O. Ilbert\altaffilmark{7},
A. Iovino\altaffilmark{14},
P. Kampczyk\altaffilmark{1},
J. Kartaltepe\altaffilmark{4},
C. Knobel\altaffilmark{1}, 
K. Kova\v{c}\altaffilmark{1},
F. Lamareille\altaffilmark{6},
J.-F. Le Borgne\altaffilmark{6},
V. Le Brun\altaffilmark{7},
M. Mignoli\altaffilmark{12}, 
Y. Peng\altaffilmark{1},
E. P\'erez-Montero\altaffilmark{6},
E. Ricciardelli\altaffilmark{9},
M. Salvato\altaffilmark{11},
J. Silverman\altaffilmark{1},
J. Surace\altaffilmark{5},
M. Tanaka\altaffilmark{8},
L. Tasca\altaffilmark{7},
L. Tresse\altaffilmark{7},
D. Vergani\altaffilmark{12},
E. Zucca\altaffilmark{12},
U. Abbas\altaffilmark{7},
D. Bottini\altaffilmark{10},
P. Capak\altaffilmark{11},
A. Cappi\altaffilmark{12},
P. Cassata\altaffilmark{7},
A. Cimatti\altaffilmark{12},
M. Elvis\altaffilmark{15},
G. Hasinger\altaffilmark{13},
A. M. Koekemoer\altaffilmark{16},
A. Leauthaud\altaffilmark{17},
D. Maccagni\altaffilmark{10},
C. Marinoni\altaffilmark{18},
H. McCracken\altaffilmark{19},
P. Memeo\altaffilmark{10},
B. Meneux\altaffilmark{7},
P. Oesch \altaffilmark{1},
R. Pell\`o\altaffilmark{6},
C. Porciani\altaffilmark{20},
L. Pozzetti\altaffilmark{12},
R. Scaramella\altaffilmark{21},
C. Scarlata\altaffilmark{5},
D. Schiminovich\altaffilmark{22},
Y. Taniguchi\altaffilmark{23},
M. Zamojski\altaffilmark{5} 
}

\altaffiltext{1}{Institute of Astronomy, Swiss Federal Institute of Technology (ETH), Wolfgang-Pauli-Strasse 27, CH-8093  Z\"urich, Switzerland.}
\altaffiltext{2}{Present address: SUPA Institute for Astronomy, University of Edinburgh, Royal Observatory, Edinburgh EH9 3HJ, Scotland, UK. E-mail address: kic@roe.ac.uk}
\altaffiltext{3}{CEA/DSM-CNRS, Universit\'e Paris Diderot, DAPNIA/SAp, Orme des Merisiers, 91191 Gif-sur-Yvette, France.}
\altaffiltext{4}{Institute for Astronomy, University of Hawaii, Honololu, HI, USA.}
\altaffiltext{5}{Spitzer Science Center. California Institute of Technology, Pasadena, CA, USA.}
\altaffiltext{6}{Laboratoire d'Astrophysique de Toulouse-Tarbes, Universit\'e de Toulouse, 14 avenue Edouard Belin, Toulouse, France.}
\altaffiltext{7}{Laboratoire d'Astrophysique de Marseille, 38 rue Fr\'ed\'eric Joliot-Curie, Marseille, France.}
\altaffiltext{8}{European Southern Observatory, Karl-Schwarzschild-Strasse 2, 85748 Garching, Germany.}
\altaffiltext{9}{Dipartimento di Astronomia, Universit\`a di Padova, Padova, Italy.}
\altaffiltext{10}{INAF-IASF Milano, via E. Bassini 15, 20133 Milano, Italy.}
\altaffiltext{11}{California Institute of Technology, Pasadena, CA, USA.}
\altaffiltext{12}{INAF Osservatorio Astronomico di Bologna, Bologna, Italy.}
\altaffiltext{13}{Max Planck Institut f\"ur Extraterrestrische Physik, Garching, Germany.}
\altaffiltext{14}{INAF Osservatorio Astronomico di Brera, via Brera 28, 20121 Milano, Italy.}
\altaffiltext{15}{Harvard Smithsonian Centre for Astrophysics, 60 Garden Street, Cambridge, MA, USA.}
\altaffiltext{16}{Space Telescope Science Institute, 3700 San Martin Drive, Baltimore, Maryland 21218, USA.}
\altaffiltext{17}{Berkeley Lab \& Berkeley Center for Cosmological Physics, University of California, Lawrence Berkeley National Lab, 1 Cyclotron Road, CA, USA.}
\altaffiltext{18}{Centre de Physique Th\'eorique, Marseille, France.}
\altaffiltext{19}{Institut d'Astrophysique de Paris, UMR 7095 CNRS, Universit\'e Pierre et Marie Curie, 98 bis Boulevard Arago, F-75014 Paris, France.}
\altaffiltext{20}{Argelander-Institut f\"ur Astronomie, Bonn, Germany.}
\altaffiltext{21}{INAF, Osservatorio di Roma, Monteporzio Catone (RM), Italy.}
\altaffiltext{22}{Department of Astronomy, Columbia University, MC 2457, 550 West 120th Street, New York, NY 10027, USA.}
\altaffiltext{23}{Ehime University, 2-5 Bunkyo-cho, Matsuyama 790-8577 Japan.}



\begin{abstract}

  We have used the zCOSMOS-bright 10k sample to identify 3244 {\em Spitzer/MIPS} \tfm-selected galaxies with $0.06< S_{24 \, \rm \mu m}\lsim 0.50 \, \rm mJy$ and $I_{\rm AB}<22.5$, over 1.5~deg$^2$ of the COSMOS field, and studied different spectral properties, depending on redshift. At $0.2<z<0.3$, we found that different reddening laws of common use in the literature explain the dust extinction properties of $\sim 80$\% of our infrared (IR) sources, within the error bars. For up to 16\% of objects, instead, the \ha/\hb ratios are too high for their IR/UV attenuations, which is probably a consequence of inhomogenous dust distributions.   In only a few of our galaxies at  $0.2<z<0.3$ the IR emission could be mainly produced by dust heated by old rather than young stars. Besides, the line ratios of $\sim 22$\% of our galaxies suggest that they might be  star-formation/nuclear-activity composite systems.   At $0.5<z<0.7$, we estimated galaxy metallicities for 301 galaxies: at least 12\% of them are securely below the upper-branch mass-metallicity trend, which is consistent with the local relation. Finally, we performed a combined analysis of the $\rm H_\delta$ equivalent-width  versus $\rm D_n(4000)$ diagram for 1722 faint and bright \tfm galaxies at $0.6<z<1.0$, spanning two decades in mid-IR  luminosity.  We found that, while secondary bursts of star formation are necessary to explain the position of the most luminous IR galaxies in that diagram, quiescent, exponentially-declining star formation histories can well reproduce the spectral properties of $\sim 40$\% of the less luminous sources.  Our results suggest a transition in the possible modes of star formation at total IR luminosities  $L_{\rm TIR} \approx (3\pm2) \times 10^{11} \, \rm L_\odot$.

\end{abstract}

\keywords
{infrared: galaxies  -- galaxies: evolution -- galaxies: abundances -- galaxies: starburst -- galaxies: active} 


\section{Introduction}
\label{sec-intro}

 Investigating the composition of the extragalactic IR background (Puget et al.~1996; Elbaz et al.~2002; Dole et al.~2006) at $z<1$ is of key importance to understand galaxy evolution and the global decline of the star formation rate density in the last 8 Gigayears (Gyr). At $z>1$, the IR background is mainly dominated by  luminous and ultra-luminous IR galaxies (LIRGs and ULIRGs, with total IR [$5-1000 \, \rm \mu m$] luminosities $10^{11}<L_{\rm TIR}<10^{12} \, \rm L_\odot$ and  $L_{\rm TIR}> 10^{12} \, \rm L_\odot$, respectively). However, these populations are progressively less important at lower redshifts, becoming rare by the present epoch. The most typical IR sources at $z \lsim 0.7$ are IR normal galaxies, i.e. those with total IR luminosities $L_{\rm TIR}< 10^{11} \, \rm L_\odot$. Their study is fundamental to clarify the production of the bulk of the IR background in the second half of cosmic time.

  Our knowledge of IR galaxies at low redshifts was first possible thanks to studies conducted with the {\em Infrared Astronomical Satellite (IRAS)} and the {\em Infrared Space Observatory (ISO)} (see e.g. Aussel et al.~1999; Genzel \& Cesarsky~2000, for a  review), which revealed that both star formation and nuclear activity were associated with the dust emission.  More recently, with the advent of the {\em Spitzer Space Telescope} (Werner et al.~2004) and the {\em Akari Telescope} (Matsuhara et al.~2006), a much deeper insight into the nature of the $z<1$ IR galaxy population has been possible. We have now characterised their luminosity evolution (e.g. Le Floc'h et al.~2005; Caputi et al.~2007; Magnelli et al.~2009), stellar masses (e.g. Caputi et al. 2006a,b), and spectral energy distributions (SEDs; e.g. Rowan-Robinson et al.~2005; Rocca-Volmerange et al.~2007; Takagi et al.~2007; Bavouzet et al.~2008; Kartaltepe et al.~2009; Symeonidis et al.~2009), among others (see Soifer et al.~2008, for a review).

 Spectroscopic surveys at different wavelengths can add very valuable information that help understand the nature of the $z<1$ IR sources. In addition to providing accurate redshift determinations, spectroscopic data allow us to characterise other aspects of IR galaxies, such as their extinction properties, metallicities and general chemical composition.  The initial optical/near-IR spectroscopic studies of IR galaxies at  $z<1$ targeted {\em IRAS} and {\em ISO} sources (e.g. Kim et al.~1995; Veilleux et al.~1995, 1999; Franceschini et al.~2003; Flores et al.~2004; Arribas et al.~2008; Garc\'{i}a-Mar\'{i}n et al.~2009). In the {\em Spitzer} era, and in the light of larger spectroscopic campaigns, more systematic studies of IR galaxies and their environments have been possible (e.g. Choi et al.~2006; Papovich et al.~2006; Caputi et al.~2008 [C08 hereafter], 2009; Desai et al.~2008).

 In particular, optical spectra are useful to disentangle the power source of the IR emission, which  generally occurs when the galaxy dust re-processes the energy provided by the ultra-violet (UV) photons of young stars or active nuclei. However, in some cases, the observed IR emission might not be associated with any on-going stellar or nuclear activity, but be rather produced by dust heated by old stars (e.g. Gordon et al.~2000; Kong et al.~2004; Cortese et al.~2006; but also see Young et al.~2009), or even direct stellar emission. The importance of these mechanisms within the IR galaxy population is not yet well established. The identification of `passive' IR sources is important to quantify the degree of contamination in the IR-derived cosmic star formation rate densities. Spectroscopic data can be very helpful to elucidate the presence of star formation in sources that are presumably dominated by old stellar populations.
 
 Spectral features also carry the imprints of the galaxy star formation history. Thus, for IR galaxies,  optical spectra can be used to investigate at what stage of the galaxy history the IR phase is produced.  Locally, the rare LIRGs and ULIRGs are known to be mainly the consequence of galaxy mergers that induce bursts of star formation and nuclear activity (e.g. Armus et al.~2007). Instead, more typical IR galaxies at $z=0$ are usually less disturbed, and appear to form their stars in a more regular way.  There are several clues that suggest that, in the past, when LIRGs and ULIRGs were more common,  both the burst-like and quiescent modes of star formation could also lead to the IR phase (e.g. Elbaz et al.~2007; Daddi et al.~2008). In spite of these different pieces of evidence, it is not yet clearly established whether there exists a transition in the possible modes of star formation between galaxies of different IR luminosities. One of the main aims of this work is to study such a transition in IR galaxies at $0.6<z<1.0$.

 The Cosmic Evolution Survey (COSMOS; Scoville et al.~2007) has been designed to probe galaxy evolution, star formation and the effects of large-scale structure up to high redshifts. The COSMOS field is defined by its {\em Hubble Space Telescope/Advanced Camera for Surveys (HST/ACS)}  2  deg$^2$ coverage (Koekemoer et al.~2007). There are multiple follow up observations carried out in the COSMOS field, ranging from X-rays to radio wavelengths. Among the photometric imaging observations, COSMOS includes the full and homogeneous coverage of the field with the {\em Spitzer Infrared Array Camera} ({\em Spitzer/IRAC}; Fazio et al.~2004) and the {\em Multiband Photometer for Spitzer} ({\em MIPS}; Rieke et al.~2004), as part of {\em Spitzer} cycle-2 and 3 Legacy Programs (Sanders et al.~2007).

 COSMOS also comprises a large spectroscopic follow-up program named zCOSMOS (Lilly et al.~2007), that is being performed with the {\em Visible Multiobject Spectrograph} ({\em VIMOS}; Le F\`evre et al.~2003) on the {\em Very Large Telescope (VLT)}. The zCOSMOS survey has two parts, one of which consists in the observation of an $I<$ 22.5 AB mag limited sample of 20,000 galaxies over 1.7 deg$^2$  of the COSMOS field  (zCOSMOS-bright).  Here, we make use of the first half of spectra observed and analysed in zCOSMOS-bright, which constitute the so-called `zCOSMOS-bright 10k sample' (Lilly et al.~2009).

  This paper presents the second part of a study we have carried out to analyse the optical spectral properties of \tfm-selected  galaxies in the COSMOS field, using the zCOSMOS-bright 10k sample.  In this second part, we analyse 3244  IR sources with $0.06 < S_{24 \, \rm \mu m} \lsim  0.50 \, \rm mJy$, which constitute the largest optical spectroscopic sample of mid-IR-selected galaxies analysed to date in the redshift range $0<z\lsim 1$. The results presented in this paper complement our first study of the optical spectra of brighter {\em MIPS} \tfm sources (C08), and focus on the most outstanding aspects discussed in this previous work. The study of these new fainter galaxies is of particular importance:  they largely outnumber our previous sample and, especially,  span the typical IR luminosities that dominate the IR background at $z<1$ (Le Floc'h et al.~2005; Caputi et al.~2007; Magnelli et al.~2009), covering a range in $L_{\rm TIR}$ of e.g. a few times $10^{10}$ to $9 \times 10^{10} \, \rm L_\odot$ at $z=0.5$, and between  $L_{\rm TIR} \approx 9 \times 10^{10}$ and $5\times 10^{11} \, \rm L_\odot$ at $z=1$.

 The layout of this paper is as follows: in Section \S\ref{sec_sample}, we describe the deep {\em MIPS} \tfm and zCOSMOS-bright datasets and the cross-correlation of the two catalogues. In Section \S\ref{sec_rflum}, we give details about the rest-frame mid-IR luminosities for our galaxies. We present our results in Sections \S\ref{sec_linelowz} and \S\ref{sec_linehighz},  based on a variety of line diagnostics in different redshift ranges: we extensively discuss the dust extinction properties of our galaxies at $0.2<z<0.3$, and study metallicities and stellar masses at $0.5<z<0.7$.  Later on, we perform a combined analysis of our present faint and previous brighter IR samples at $0.6<z<1.0$ to investigate the possible mechanisms of star formation in IR galaxies, and how they vary with IR luminosity.  Finally, we summarise and discuss our results in Section \S\ref{sec_concl}.  We adopt throughout a cosmology with $\rm H_0=70 \,{\rm km \, s^{-1} Mpc^{-1}}$, $\rm \Omega_M=0.3$ and $\rm \Omega_\Lambda=0.7$.   All quoted magnitudes and colours refer to the AB system.

\section{Datasets}
\label{sec_sample}
 
\subsection{The zCOSMOS-bright 10k sample} 

 The zCOSMOS survey (Lilly et al.~2007) is a very large spectroscopic program  being performed with {\em VIMOS}~(Le F\`evre et al.~2003) on the {\em VLT}.  {\em VIMOS} is a  multi-slit spectrograph with four non-contiguous quadrants that cover 7$\times$8 arcmin$^2$ each. zCOSMOS is designed to have a uniform pattern of pointings and a multiple-pass strategy, which guarantee a uniform coverage of the field. The technical characteristics of this survey are explained in further detail by Lilly et al.~(2007).
 
 In zCOSMOS-bright, the vast majority of targets are selected randomly from a complete $I< 22.5$ AB mag catalogue (the `parent catalogue'). This selection $I$-band magnitude corresponds to a total {\em HST/ACS} F814W magnitude in most cases. zCOSMOS-bright also includes a set of targets that have been selected for specific reasons, e.g. X-ray or radio sources. All the observations are performed with the $R \sim 600$ {\em VIMOS} medium-resolution (MR) grism, with a spectral coverage of $(5500-9500) \, \rm \AA$ and a dispersion of $2.55 \, \rm \AA$. The spectra are automatically reduced using the VIPGI software (Scodeggio et al.~2005), and the redshift determination of each source is manually checked  on an individual basis by two independent reducers.

  The confidence in the redshift determination of each source is qualified with a flag, whose values can be: 4 (completely secure redshift); 3 (very secure redshift, but with a very marginal possibility of error); 2 (a likely redshift, but with a significant possibility of error); 1 (possible redshift); 0 (no redshift determination); and 9 (redshift based on a single secure narrow line, which is usually \oii or \ha). The addition of $+10$ to the flag indicates a broad-line active galactic nucleus (BLAGN). The confidence flags are manually assigned and reconciled by the two independent reducers (see Lilly et al.~2007, 2009 for further details). 
  
 Independently, photometric redshifts for the entire zCOSMOS sample have been obtained with The Zurich Extragalactic Bayesian Redshift Analyzer  (ZEBRA; Feldmann et al.~2006), using the UV through near-IR waveband data available for the COSMOS field.  Also, a newer set of photometric redshifts based on 30-band photometry have been derived by Ilbert et al.~(2009) and Salvato et al.~(2009).  Tests performed on objects with duplicate spectra of different quality show that good-quality photometric redshifts as those obtained with the COSMOS datasets can be useful to confirm less-secure spectroscopic redshifts, such as those with flag=2, 1 or 9 (see Lilly et al.~2007). The differences between the photometric and spectroscopic redshifts are $\Delta z<0.08 \times (1+z)$ for $\geq 93$\% of the objects with spectroscopic flag=2, 3 or 4, and for 72\% of those with a flag=1 (Lilly et al.~2009).  
  
 In this paper, we make use of the `zCOSMOS-bright 10k sample' over 1.5 deg$^2$ of the COSMOS field, in which 10,580 out of 10,644 sources have  $I< 22.5$ AB mag. This first set of zCOSMOS-bright spectra are now publicly available\footnote{http://archive.eso.org/cms/eso-data/data-packages/zcosmos-data-release-dr2/}. A full description of this dataset can be found in  Lilly et al.~(2009).

\subsection{The SCOSMOS MIPS deep \tfm catalogue}

 The SCOSMOS survey (Sanders et al.~2007) comprises {\em IRAC} 3.6, 4.5, 5.8 and 8.0 $\rm \,\mu m$ and {\em MIPS} 24, 70 and 160 $\rm \,\mu m$ observations including the entire 2 deg$^2$ of the COSMOS-ACS field, and are part of the {\em Spitzer} cycle-2 and 3 Legacy Programs. 
  
 In cycle-3, the COSMOS field has been mapped at \tfm down to a flux density $S_{24 \, \rm \mu m}\approx 0.06 \, \rm mJy$ (SCOSMOS-deep). The source extraction and photometric measurements on the \tfm maps have been performed with the SExtractor software (Bertin \& Arnouts~1996) and the DAOPHOT package (Stetson~1987), respectively. A point spread function (PSF) fitting technique for measuring the photometry has been  necessary to deal with blending problems. Further details on the SCOSMOS survey as well as the data reduction and source detection procedures are given by Sanders et al.~(2007) and Le Floc'h et al.~(2009). 
 
\begin{figure}
\epsscale{0.7}
\plotone{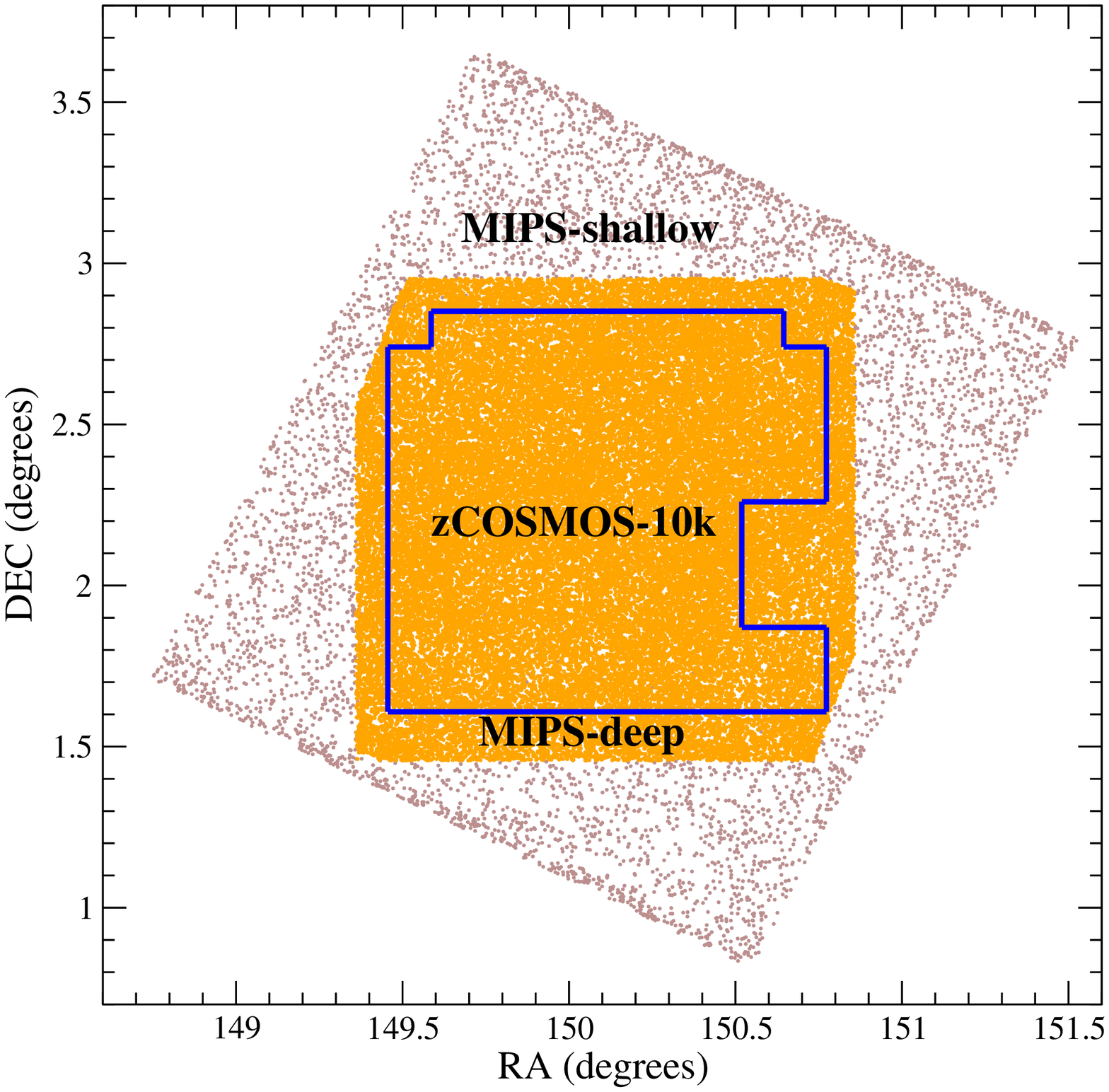}
\caption[]{\label{fig_field} The zCOSMOS-bright 10k and the SCOSMOS/MIPS shallow and deep sample coverage fields.}
\end{figure}

 The final SCOSMOS-deep \tfm catalogue contains 52,092  sources with $S_{24 \, \rm \mu m}> 0.06 \, \rm mJy$  over the entire COSMOS field, with $\sim$70\% of them lying within the zCOSMOS-bright 10k-sample area (see Figure \ref{fig_field}).

\subsection{Cross-correlation of the catalogues}

We performed the cross-correlation of the {\em MIPS}-deep and zCOSMOS-bright 10k catalogues in the same way as we did it for the shallow {\em MIPS} \tfm sample in C08, so we only summarise the procedure here, quoting the figures corresponding to the matching of the deep \tfm sample.

Firstly, as the vast majority of our zCOSMOS-bright 10k sources are randomly taken from an  $I< 22.5$ AB mag complete catalogue, it is important to make clear what fraction of the  $S_{24 \, \rm \mu m}> 0.06 \, \rm mJy$ sources can be identified with such an optical magnitude cut. To determine this fraction, we cross-correlated the SCOSMOS-deep \tfm catalogue with the zCOSMOS-bright `parent catalogue'. We obtained that around 39\% of  $S_{24 \, \rm \mu m}> 0.06 \, \rm mJy$ sources are associated with an $I< 22.5$ AB mag counterpart and, therefore, our analysis can be considered as representative of this fraction of \tfm sources.

The cross-correlation of  the deep \tfm catalogue with the zCOSMOS-bright 10k catalogue allowed us to identify 4497 IR sources, using a matching radius of 2 arcsec. This figure is the result of simultaneously considering: the fraction of all \tfm sources lying in the zCOSMOS-bright 10k-sample field ($\sim$70\%); the fraction of sources selected with the $I< 22.5$ mag cut (39\%), and the zCOSMOS-bright 10k sampling rate, i.e. $\sim 1/3$ of the $I< 22.5$ sources.

To assess the presence of false identifications, we performed a separate cross-correlation of the deep \tfm catalogue with a deep complete  $I<25$ AB mag  catalogue available for the COSMOS field (Capak et al.~2007). We found that a small fraction of our \tfm sources (138 out of 4497, i.e. 3\%) have an optical $I<25$ AB mag association that is closer than the zCOSMOS-bright 10k counterpart. We excluded these 138 sources from our analyisis.

Out of the remaining 4359 \tfm/zCOSMOS identifications,  3865 sources (i.e. $\sim$89\%) have a good-quality redshift determination (as indicated by their zCOSMOS quality flags; see Lilly et al.~2007 and C08).  3838 of these sources are galaxies, while the other 27 are galactic stars.

Finally, our sample of 3838 \tfm galaxies with good-quality zCOSMOS-bright redshifts and spectra contains 594 objects that we have already analysed in C08. (Note that the very small discrepancy of 2\% in the number of bright \tfm galaxies analysed in C08 and found here is due to the update of the \tfm and zCOSMOS-bright 10k catalogues). In this work, we study new 3244 galaxies with $0.06 \, \rm mJy < S_{24 \, \rm \mu m} \lsim 0.50 \, \rm mJy$, and combine or compare with our previous results for the brighter \tfm sample, where appropriate. 

\begin{figure}
\epsscale{0.8}
\plotone{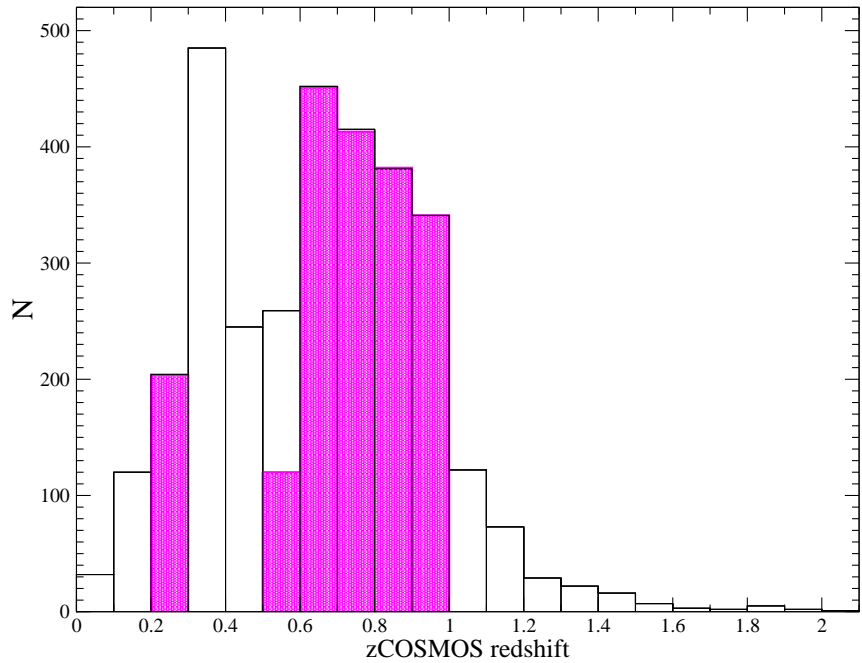}
\caption[]{\label{fig_zhisto} The zCOSMOS redshift distributions of the new \tfm sources identified in this work, and those considered for spectroscopic analysis (empty and filled histograms, respectively).}
\end{figure}

Nearly 97 (99)\% of the 3244 galaxies in our final sample lie at $z<1.2$ ($z<1.5$), as it can be seen in Figure \ref{fig_zhisto}. This figure also indicates the redshift bins and subset of galaxies that we effectively consider for spectroscopic analysis in the following sections.   The 1\% high-redshift tail at $z \geq 1.5$ is almost exclusively composed of sources identified as BLAGN with the zCOSMOS spectra. Within our \tfm/zCOSMOS sample, there are 88 BLAGN in total at any redshift. Roughly half of them are at $z \geq 1.5$.

\section{IR luminosities}
\label{sec_rflum}

 We computed the rest-frame \tfm luminosities \nLntfm  of the 3244 galaxies in our final sample in the same way as in C08: we used the observed  \tfm fluxes, the zCOSMOS redshifts, and k-corrections derived using the Lagache et al.~(2003, 2004) IR spectral energy distribution (SED) models.  Our galaxies have \tfm luminosities $\nu L_\nu^{\rm 24 \, \mu m} \approx 10^8 \, L_\odot$ at $z=0.1$, roughly between  $10^9$ and $10^{10} \, L_\odot$ at $z=0.5$, and between  $10^{10}$ and $\sim 10^{11} \, L_\odot$ at $z=1$.

 The estimated  total IR luminosities $L_{\rm TIR}$ of these galaxies depend on the recipe used to convert monochromatic  into bolometric IR luminosities. By adopting the \nLntfm-$L_{\rm TIR}$ calibration of Bavouzet et al.~(2008), we obtain that our galaxies span $L_{\rm TIR}$ values between a few times $10^{10}$ and   $9 \times 10^{10} \, L_\odot$ at $z=0.5$, and  $L_{\rm TIR} \approx 9 \times 10^{10}$ to $5 \times 10^{11} \, \rm L_\odot$ at $z=1$. These total IR luminosities imply that our sample is composed of IR normal galaxies at intermediate redshifts, and mainly LIRGs by $z=1$.   The analysis of IR sources in these luminosity ranges is particularly important, as they make the bulk of the IR background at these redshifts (Le Floc'h et al.~2005; Caputi et al.~2007; Magnelli et al.~2009). 

In general, the SED models and recipes used to estimate IR luminosities apply only to IR galaxies dominated by star formation. The SED of some  AGN, instead, are characterised by a power-law shape $f_{\nu} \propto \nu^\alpha$  (with $\alpha<0$; see e.g. Elvis et al.~1994; Alonso-Herrero et al.~2006). As in C08, we analysed the {\em Spitzer/IRAC} photometry available for our galaxies to determine those cases in which the fluxes in the four {\em IRAC} channels 3.6, 4.5, 5.8 and 8.0 $\rm \mu m$ are consistent  with a single power law, within the error bars. We found that 96 of our 3244 galaxies have an  {\em IRAC} power-law SED, and nearly 80\% of them are also zCOSMOS-classified BLAGN.

\section{Spectral diagnostics at $0.2<\lowercase{z}<0.3$}
\label{sec_linelowz}

\subsection{Balmer lines and optical colours}
\label{sec_balcol}

The wavelength coverage (5500-9500)~$\rm \AA$ of the {\em VIMOS} MR zCOSMOS spectra  allows for different Balmer lines to be observed at different redshifts. The spectral quality is quite uniform  up to wavelengths $\lambda \approx 8000 \rm \AA$, but is degraded by fringing at longer wavelengths, especially at $\lambda \gsim 8500 \rm \AA$. Thus, for example, although one can  in principle observe the \ha line up to redshift $z\approx 0.45$, reliable measurements of this line on individual spectra are generally only possible for sources at $z\lsim 0.3$.

We measured the fluxes and equivalent widths (EWs) of all lines in our spectra by direct integration on the rest-frame spectra. To minimise the errors introduced by fringing, we manually checked the measurements of those lines at observed  $8000 < \lambda < 8500 \, \rm \AA$ which had particular problems, such as a significantly shifted centre. We have not performed line measurements beyond observed $\lambda \gsim 8500 \rm \AA$. 

In this section, we focus on the analysis of those spectra in which the two main Balmer lines \ha and \hb are simultaneously present, as their flux ratio provides information on the dust obscuration affecting the on-going star formation in IR galaxies. With this purpose, and willing to avoid measuring lines both in the fringing region and the blue edge of the spectra, we performed simultaneous \ha and \hb measurements only for the 204 IR galaxies that lie at $0.2<z<0.3$ within our sample (none of which is classified as a BLAGN).

 We corrected both the \ha and \hb emission line measurements for stellar absorption. We estimated the stellar absorption correction in each case by fitting the continuum of each spectrum with synthetic stellar SEDs from the Bruzual \& Charlot 2007 library (Bruzual \& Charlot~2003; Bruzual~2007). We also applied aperture corrections to account for the fact that the slits in the {\em VIMOS} masks have a width of only 1 arcsec. To obtain the aperture corrections, we convolved each zCOSMOS spectrum in our sample with the {\em Subaru Suprime-Cam} $R$ and $I$-band filters, and compared the resulting magnitudes with the total magnitudes of our galaxies, as available in the COSMOS optical photometry catalogues (Capak et al.~2007).

 Within our sample at $0.2<z<0.3$,  189 out of 204 ($>$92\%) galaxies  have \ha $\rm EW > 5 \, \AA$ (before stellar absorption correction). This means that these galaxies have an on-going star formation that is significant with respect to the underlying stellar population.  The remaining 15 galaxies with \ha $\rm EW < 5 \, \AA$ will be analysed later, but we anticipate that only among them can we expect to have sources in which the IR emission is likely to be associated with dust heated by old stars rather than UV photons from on-going star formation. The adopted limit ($\rm EW = 5 \, \AA$) to separate our galaxy sample is the same as the one we used in C08 and, of course, is arbitrary. However, no result or conclusion in this work depends on the exact value of that limit.

Among the 189 \tfm  galaxies with \ha $\rm EW > 5 \, \rm \AA$ at $0.2<z<0.3$, 94 also have \hb $\rm EW > 5 \, \rm \AA$, while the other 95 have lower \hb EW values. This is not very surprising as, even in the absence of extinction, there is no reason for both \ha and \hb EWs to be similar. Their ratio depends on the galaxy age and star formation history (because they determine the slope of the continuum between  the \ha and \hb wavelengths). In addition, any possible differential extinction between gas and stars could also influence the EW ratio.  We discuss in more detail the extinction properties of these sources in  Section \ref{sec_dust}.

\begin{figure}
\epsscale{1.0}
\plotone{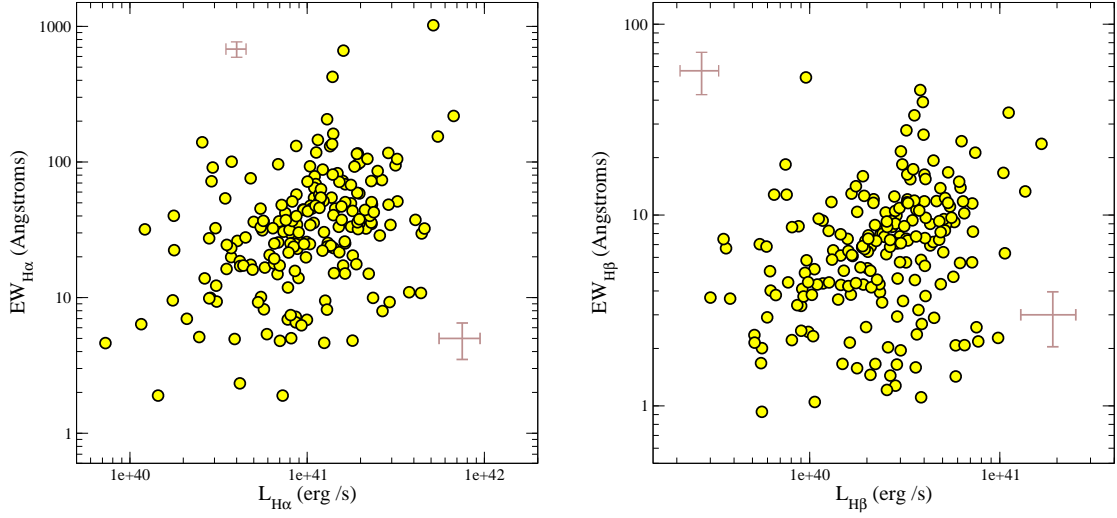}
\caption[]{\label{fig_ewlum} \ha and \hb EWs versus luminosities for our IR galaxies at $0.2<z<0.3$ (left and right-hand panels, respectively). All measurements are corrected for stellar absorption, and line luminosities also include aperture corrections.}
\end{figure}

Although here we classify our galaxies according to their \ha and \hb EWs (which measure the relative importance of the on-going and the past integrated star formation histories), it is instructive to see the behaviour of these quantities with respect to the line luminosities (Figure \ref{fig_ewlum}). Globally, the EWs and luminosities relate to each other, although with non-negligible dispersion. This is both due to the different optical continuum levels in IR galaxies, and the spectral measurement errors. 

 In this section, our main interest is to investigate whether IR galaxies with different Balmer line EWs can be recognised by their optical broad-band colours, and how, in general, the optical colour distribution of IR galaxies compares with the colour distribution of the entire zCOSMOS 10k sample. 
 
\begin{figure}
\epsscale{1.0}
\plotone{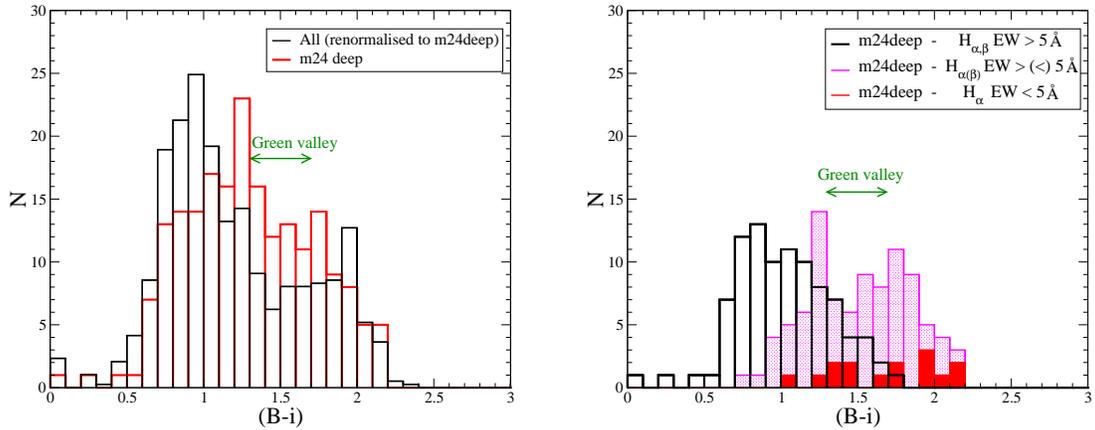}
\caption[]{\label{fig_bmihisto} Left: the distribution of observed $(B-i)$ colours for \tfm-detected and all zCOSMOS galaxies at $0.2<z<0.3$. Right: the $(B-i)$ colour distributions for the \tfm galaxies segregated by their Balmer-line EWs.}
\end{figure}

 The left-hand panel of figure \ref{fig_bmihisto} shows the distribution of the observed $(B-i)$ colours for all zCOSMOS-bright 10k galaxies at $0.2<z<0.3$, compared to the colour distribution for the \tfm/zCOSMOS sources (note that the number of all zCOSMOS-bright galaxies is renormalised to 204 to coincide with the number of \tfm/zCOSMOS sources in this redshift range). On the right-hand panel, we plot the respective $(B-i)$  colour distributions for the \tfm/zCOSMOS sources with  \ha and \hb $\rm EW > 5 \, \AA$; \ha  $\rm EW > 5 \, \AA$ and \hb  $\rm EW < 5 \, \AA$; and both $\rm EW < 5 \, \AA$.  In all cases, the $(B-i)$ are total (i.e. aperture-corrected) colours measured on the {\em Subaru Suprime-Cam} images for COSMOS (Taniguchi et al.~2007).

 The comparison of these different histograms clearly shows that, while the overall zCOSMOS galaxy population at $0.2<z<0.3$ displays a bimodal colour distribution, this bimodality is not seen for the IR sources. This is basically due to the relatively higher  fraction of `green valley' $1.3 \leq (B-i) \leq 1.7$ galaxies within the \tfm sample, and lower fraction of blue $(B-i)<1.3$ galaxies, which suggests that the zCOSMOS sources detected at \tfm are on average more obscured.

When we analyse the  $(B-i)$ colour distributions for the IR galaxy samples with different values of \ha and \hb EWs, we see the following:  those galaxies with both \ha and \hb $\rm EW > 5 \, \AA$ make most ($\sim$ 75\%) of the optically-blue $(B-i)<1.3$ IR galaxies. Instead, the `green valley' and red IR galaxies are mainly  sources with  \hb $\rm EW < 5 \, \AA$  (58\% and 78\%, respectively). Therefore, although there is no one-to-one relation between the line EWs and the broad-band flux ratios, one can roughly predict the spectral characteristics of IR galaxies according to their different optical colours.

\begin{figure}
\epsscale{0.70}
\plotone{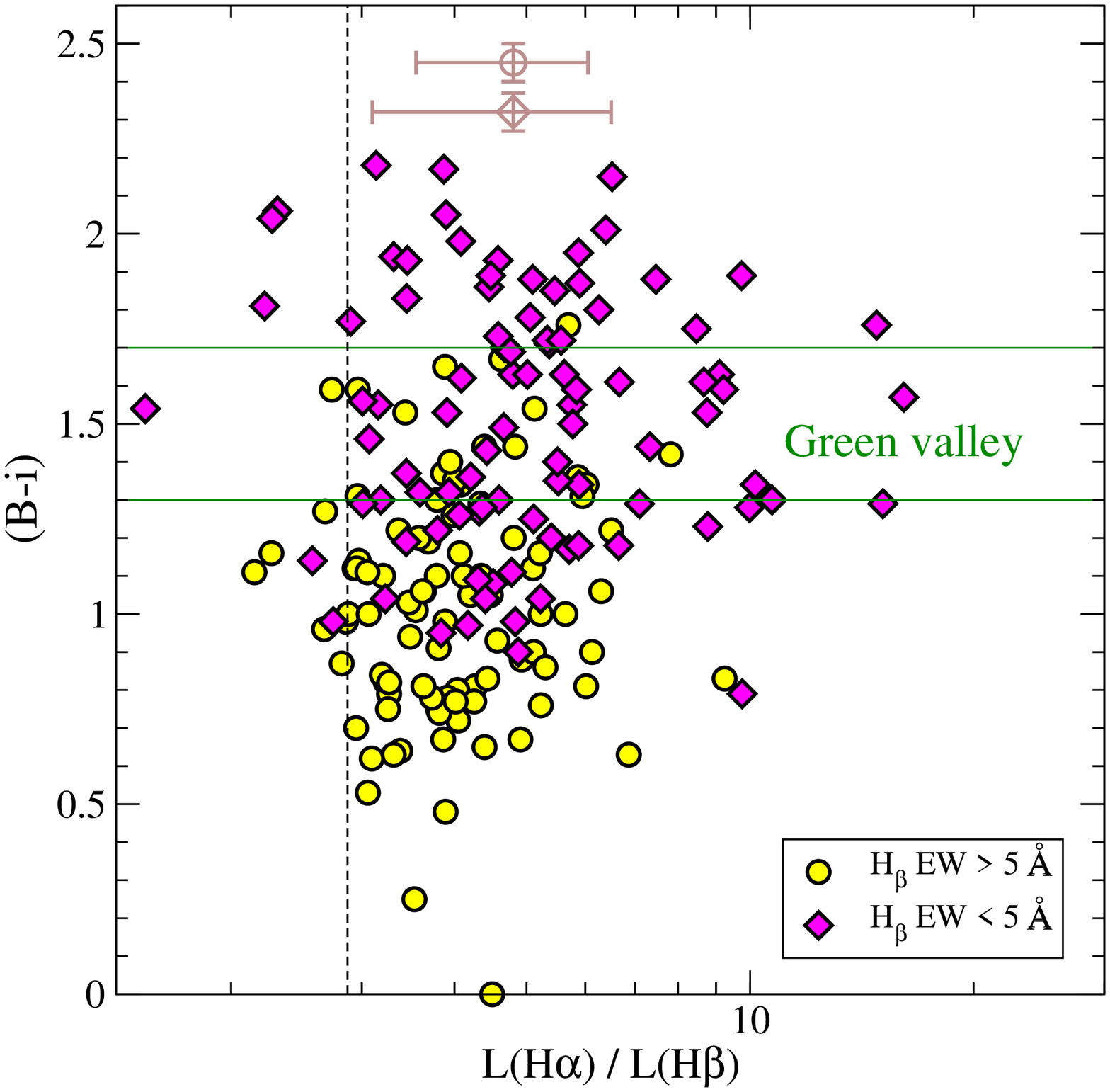}
\caption[]{\label{fig_bmivshahb} The observed $(B-i)$ colour versus the Balmer decrement for our  \tfm galaxies with \ha $\rm EW > 5 \, \AA$ at $0.2<z<0.3$. Typical error bars for galaxies with \hb $\rm EW >$ and  $\rm < 5 \, \AA$ are shown at the top of the plot (with a circle and a diamond, respectively). The vertical dashed line indicates the \ha/\hb ratio expected in a case B recombination with $T=10,000 \, \rm K$ (Osterbrock~1989).}
\end{figure}

A similar conclusion can be extracted by plotting the observed $(B-i)$ colours as a function of the Balmer decrement  $L(H\alpha)/L(H\beta)$ (Figure \ref{fig_bmivshahb}). In this diagram, it is clear that galaxies with \hb $\rm EW < 5 \, \AA$ dominate the region defined by $(B-i) \geq 1.3$. The vast majority of our \tfm galaxies  display \ha/\hb values above the  intrinsic ratio in the typical case B recombination with $T=10,000 \, \rm K$ (i.e. \ha/\hb$=$2.87; Osterbrock 1989),  independently of their optical colours, indicating in all cases significant dust attenuations. However, nearly all the sources with the highest dust extinctions \ha/\hb$>7$  have $(B-i) \gsim 1.3$ (cf. Section \S\ref{sec_dust}).

\begin{figure}
\epsscale{0.60}
\plotone{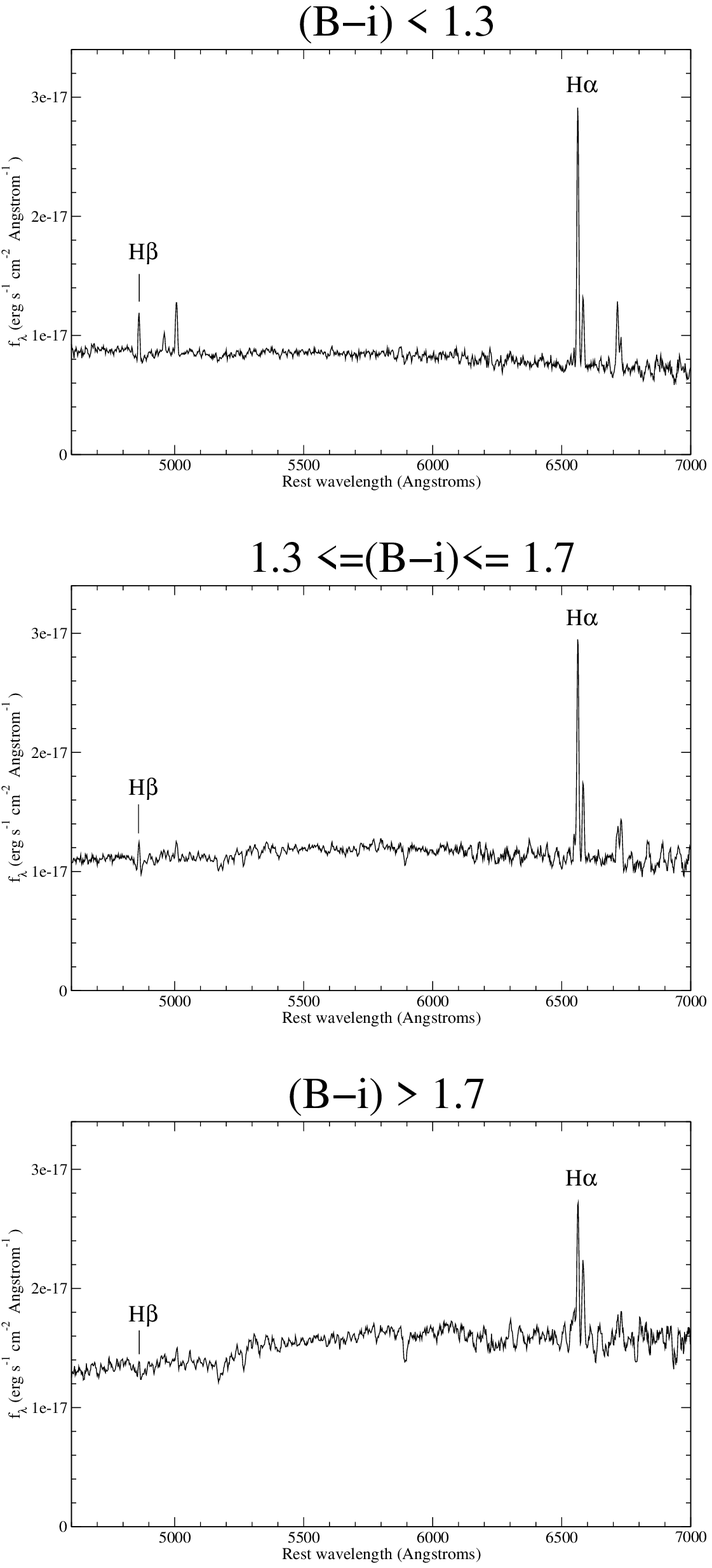}
\caption[]{\label{fig_stackhanohb} The average stacked spectra of  $0.2<z<0.3$ \tfm galaxies with individual \ha $\rm EW > 5 \, \AA$ and \hb $\rm EW < 5 \, \AA$, according to their $(B-i)$ colours. }
\end{figure}

The stacked spectra of the sources with  \hb $\rm EW < 5 \, \AA$, classified according to their $(B-i)$ colours, are useful to summarise the average spectral properties (Figure \ref{fig_stackhanohb}). These stacks show that the Balmer line EWs become smaller for redder colours, with the (uncorrected) \hb EW ranging from $\rm (3.48\pm0.50) \, \AA$ for the $(B-i)<1.3$ stack, to $\rm (0.62\pm0.20) \,  \AA$ for that with $(B-i)>1.7$. The average  stellar-absorption-corrected $L(H\alpha)/L(H\beta)$ ratio varies from $(4.12 \pm 0.52)$ to $(10.3\pm2.6)$, respectively.

\subsection{IR galaxies with H$\alpha$ EW $< 5\, \rm \AA$}

Our \tfm/zCOSMOS galaxy sample at $0.2<z<0.3$ contains 15 sources whose individual \ha  EWs are smaller than $5 \, \rm \AA$. This indicates that either the on-going star formation in these galaxies is insignificant with respect to their integrated past star formation history, or that there is a very different degree of extinction in the gas associated with the new star formation and the old stars responsible for the continuum light.  In this section, we investigate in detail the properties of these galaxies, and discuss in particular the origin of their IR emission.

\begin{figure}
\epsscale{0.60}
\plotone{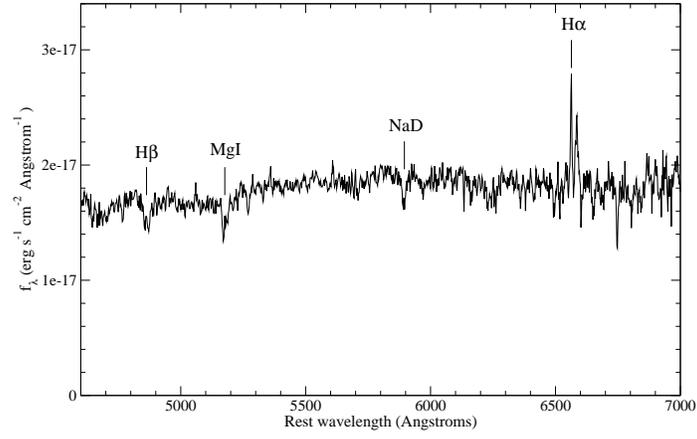}
\caption[]{\label{fig_stacknoha} The average stacked spectrum of  \tfm galaxies at  $0.2<z<0.3$ with individual \ha $\rm EW < 5 \, \AA$. }
\end{figure}

Figure \ref{fig_stacknoha} shows the average stacked zCOSMOS spectrum of these 15 galaxies.  This spectrum is characterised by a weak \ha line in emission with $\rm EW=(4.8\pm1.0) \, \AA$, and an \hb line in absorption that indicates the importance of the underlying old stellar populations. The presence of prominent absorption MgI line at $\lambda=5175.4 \, \rm \AA$  and NaD line at $\lambda=5892.5 \, \rm \AA$  confirms the dominance of the old stellar component.

To better understand the nature of our \ha $\rm EW < 5 \, \AA$ sources, we inspect their {\em HST/ACS} F814W-band morphologies on an individual basis. The {\em ACS} images reveal a variety of cases: some of these galaxies are irregular with clumps, favouring the case of heavily obscured star formation. A few are simply edge-on galaxies, so the large extinction is mainly an orientation effect. 

Five out of 15 of these sources are ellipticals or lenticular galaxies. In principle, there is no indication of significant star formation in these objects except for their IR emission. Only two of them are detected in the near-UV with $NUV \lsim 25.8$ mag (cf. Section \ref{sec_dust}). These galaxies are probably the only candidates in our $0.2<z<0.3$ sample for which the IR emission could be mainly associated with dust heated by the UV photons of evolved stars. 

\begin{figure}
\epsscale{0.80}
\plotone{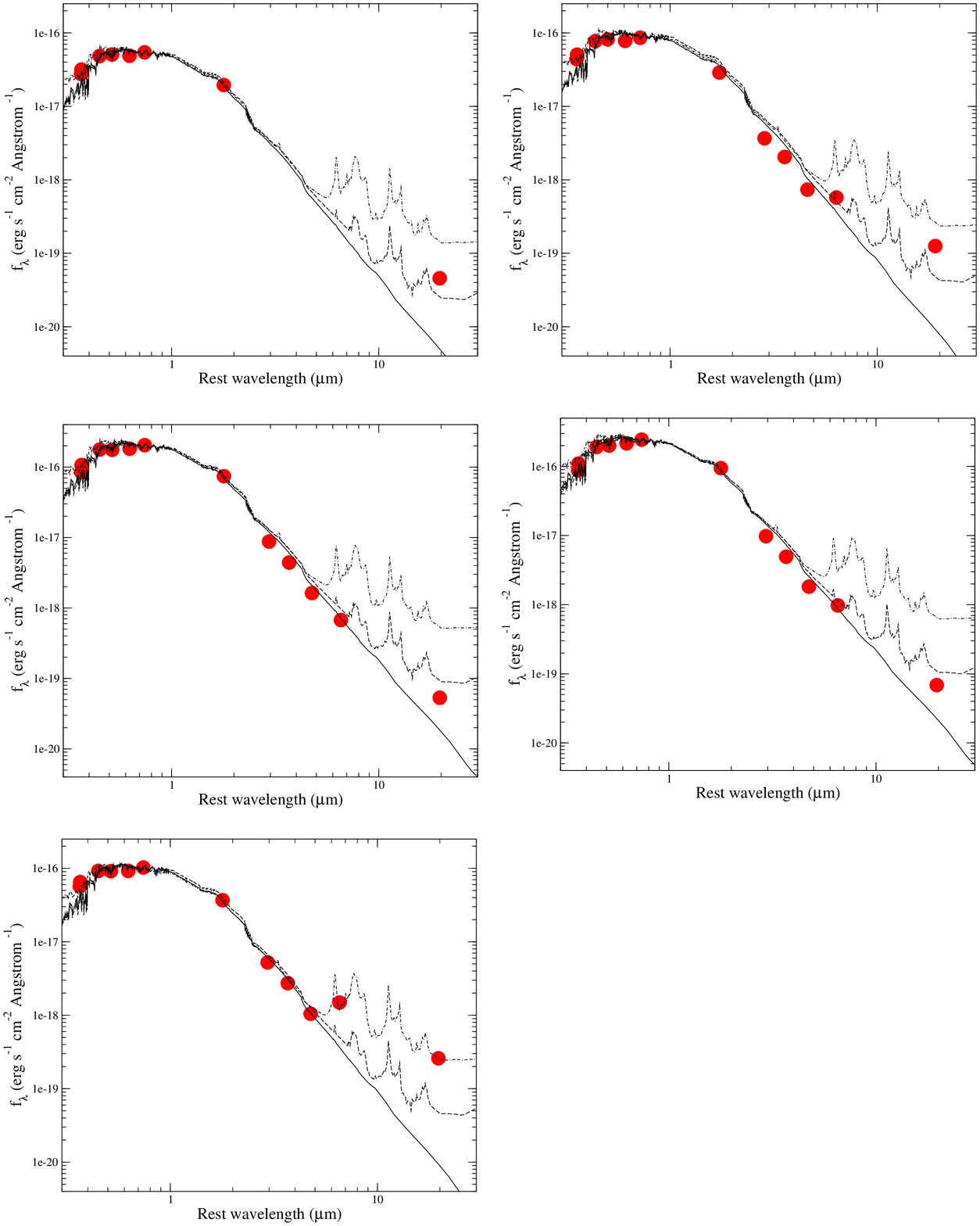}
\caption[]{\label{fig_sednoha} Broad-band SEDs of galaxy candidates for having dust heated by old rather than young stars ($0.2<z<0.3$). Circles indicate rest-frame total broad-band fluxes (note that the error bars are smaller than the symbol size). Lines of different style correspond to renormalised SED model templates from Polletta et al.~(2007), constructed using the GRASIL code (Silva et al.~1998) and empirical IR spectra: Elliptical (solid), S0 (dashed) and Sc (dot-dashed). }
\end{figure}

Figure \ref{fig_sednoha} shows the rest-frame broad-band SEDs of these five galaxies with presumably no star formation. For reference, we added the SED model templates of three galaxy types: Elliptical, S0 and Sc (from Polletta et al.~2007; based on the GRASIL code, Silva et al.~1998). For the spiral models, the templates in the wavelength range 5-12$\, \rm \mu m$ have been constructed using empirical spectra observed with {\em ISO} and {\em Spitzer} (see Polletta et al.~2007 for more details).

The comparison of our galaxy SEDs with the different models shows that, in all cases, the mid-IR emission is in excess of what is expected from pure stellar emission (as in the case of an elliptical galaxy model). This indicates that the presence of dust is indeed necessary to explain the mid-IR emission in our galaxies. Note, however, that in the two middle-panel SEDs the mid-IR excess is small, i.e. the direct stellar emission constitutes a significant part of the \tfm flux.

The IR emission of none of these five sources  seems to be powered by an AGN, as none of them has an {\em IRAC} power-law SED or is detected in the X-ray catalogues available for the COSMOS fields (Hasinger et al.~2007; Elvis et al.~2009). The mid-IR luminosity distribution of these galaxies is similar to that of IR sources with larger-EW Balmer lines at similar redshifts.

\subsection{Dust extinction}
\label{sec_dust}

As in C08, we analyse our \tfm galaxy sample with simultaneous \ha and \hb \linebreak measurements at $0.2<z<0.3$  to investigate the relation between the Balmer decrement $L(H\alpha)/L(H\beta)$  and the far-IR-to-ultra-violet (UV) luminosity ratio.  In general, for star forming galaxies, the far-IR/UV ratio gives a measurement of the dust extinction $A_{2000}$:  $(L_{\rm 70 \, \mu m}+L_{2000})/L_{\rm 2000} \approx 10^{0.4\,A_{2000}}$ (e.g. Buat \& Xu~1996; Gordon et al.~2000; Buat et al.~2002).

Our results for bright IR galaxies  with \ha and \hb EW $>5 \rm \, \AA$ at $0.2<z<0.3$ (C08) showed that, for most of them, the Balmer decrement versus the far-IR/UV attenuation was compatible with the Calzetti et al.~(2000) reddening law, within the error bars. Here, we would like to investigate whether this conclusion is still valid when considering fainter IR galaxies.

We estimated the rest-frame 70 $\rm \mu m$ luminosity of our \tfm galaxies at $0.2<z<0.3$ using their rest-frame \tfm luminosities and the  \nLntfm-$\nu L_\nu^{\rm 70 \, \mu m} \,$ conversion formula calibrated by Bavouzet et al.~(2008). To obtain UV luminosities at rest-frame $\lambda=2000 \,\rm \AA$, we used the GALEX near-UV data available for the COSMOS field (P.I.: D. Schiminovich), which has an observed effective wavelength $\lambda_{\rm eff.}=2310 \, \rm \AA$. For NUV non-detected sources, we adopted a lower limit of 25.8 AB mag.

\begin{figure}
\epsscale{0.90}
\plotone{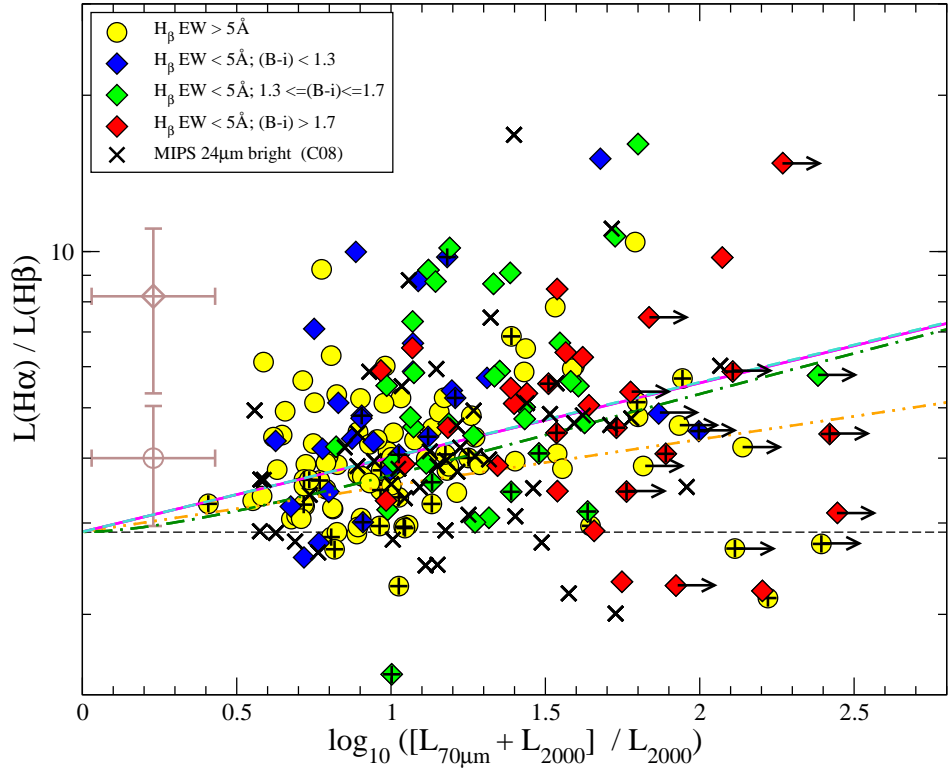}
\caption[]{\label{fig_extinct} The Balmer decrement versus the far-IR/UV attenuation.  Symbols are explained in the plot label. Right-pointing arrows indicate lower limits for objects non-detected in the NUV. Plus-like symbols within the circles/diamonds indicate objects that are in the composite region of the BPT diagram (see Figure \ref{fig_bpt}).  Typical error bars are shown on the left-hand side of the plot. Lines of different styles indicate the relation predicted with different reddening laws: SMC (Pr\'evot et al.~1984; double-dot-dashed); MW (Fitzpatrick~1999; dashed); Calzetti et al.(~2000; solid) and a composite law (dot-dashed). The horizontal dashed line shows the \ha/\hb ratio expected in a case B recombination with $T=10,000 \, \rm K$ (Osterbrock~1989). }
\end{figure}

Figure \ref{fig_extinct} shows the Balmer decrement $L(H\alpha)/L(H\beta)$ versus the extinction $\log_{10} ([L_{\rm 70 \, \mu m}+L_{2000}]/L_{\rm 2000})$, for our galaxies with \ha  EW $>5 \rm \, \AA$ at $0.2<z<0.3$ (circles and diamonds for \hb EW $>$ and $<5 \rm \, \AA$, respectively).  The values obtained for brighter IR galaxies are also shown (crosses).  Note that Figure \ref{fig_extinct} shows more bright \tfm galaxies than those analysed in C08, as here we want to study all galaxies with \ha EW $>5 \rm \, \AA$ (independently of the \hb EW), as we do for our current fainter sample. This allows us to extend our dust extinction analysis to the most obscured sources.

We see that, when considering all bright and faint \tfm galaxies with \ha EW $>5 \rm \, \AA$, the spread in the extinction diagram is quite large. While many of our sources lie  above or close to the relation derived from the Calzetti et al.~(2000) reddening law, several others have lower  Balmer decrements for their $L_{\rm IR}/L_{\rm UV}$ attenuations. The  attenuation of some of these galaxies seems more compatible with the Small Magellanic Cloud reddening law (SMC; Pr\'evot et al.~1984), but unfortunately it is difficult in general to decide among the different reddening laws given the line-ratio error bars.

Around 5\% of our \tfm galaxies are characterised by a $L(H\alpha)/L(H\beta)$  ratio close to a $T=10,000 \, \rm K$ case B recombination, but have a substantial $L_{\rm IR}/L_{\rm UV}$ attenuation. This effect has also been discussed in C08. These are usually cases of sources with an inhomogeneous dust distribution, where the {\em VIMOS} 1-arcsec slit only receives the optical light of the practically unobscured central region, while the  large ($\approx$6-arcsec full-width half maximum) {\em Spitzer/MIPS} point spread function collects the IR light of the dustier surroundings. However, in our new fainter sample, we identify a more interesting property for several of these outlier galaxies: their line ratios suggest that they could be composite star-forming/AGN systems (see Section \S\ref{sec_bpt}). For these sources, trying to understand their dust attenuation with reddening laws calibrated on pure star-forming galaxies is probably misleading.

On the other hand,  we see that other galaxies in our sample are characterised by very large Balmer decrements: around 16\% of them are at least 1$\rm \sigma$ away from any of the common reddening laws (this figure includes $\sim$13 and 19\% of sources with \hb EW $>5 \rm \, \AA$ and $<5 \rm \, \AA$, respectively). The use of any of these reddening laws with the corresponding  $L_{\rm IR}/L_{\rm UV}$ attenuations would significantly under-estimate the  \hb extinctions observed in these galaxies.  Interestingly, very few of them have line ratios characteristic of composite systems (\S\ref{sec_bpt}), so the plausible presence of obscured nuclear activity does not appear to be related to the high Balmer decrements.

Among the brighter IR galaxies in the C08 sample, and taking into account all the sources with  \ha EW $>5 \rm \, \AA$, we find that around $\sim$11\% have  Balmer decrements that are significantly above the predictions of the various reddening laws. 

A possible caveat is that these outliers could be simply produced by an under-estimation of the stellar absorption corrections applied to the \hb lines, whose median is $(3.1\pm0.3) \, \rm \AA$ and maximum value is $3.7 \, \rm \AA$. Such a correction is comparable to the same un-corrected \hb EWs in some cases. To test this possibility, we calculated the hypothetical stellar absorption corrections that would be needed  in order to make the outliers have Balmer ratios within 1$\rm \sigma$ of the Calzetti et al.~(2000) reddening law.  For this to happen, we found that around a half of the outliers would need stellar absorption corrections of more than $3.7 \, \rm \AA$, the maximum value obtained for any of our galaxies by fitting their spectral continuum. This fact suggests that, at least for half of them, the large displayed Balmer decrements  are not an artifact of under-estimated stellar absorptions. Note that, in this exercise, we have not considered any variation in the \ha stellar absorption  (in reality, the two corrections would increase simultaneously). Thus, our test determines the maximum possible fraction of outliers that could be corrected just by applying a larger, though still realistic, stellar absorption correction.

A likely explanation for the highly extincted galaxies  is that they contain  heavily obscured starbursts which are concentrated in their central regions. While  $(L_{\rm 70 \, \mu m}+L_{2000})/L_{\rm 2000}$  is related to the total amount of galaxy dust, $L(H\alpha)/L(H\beta)$ is more sensitive to the dust concentration. So, even if the total amount of dust is not very large, the \ha/\hb ratio will be,  when the dust is too concentrated towards the centre (the \ha and \hb fluxes correspond to the central 1-arcsec width light collected in the {\em VIMOS} slit). Note that basically all of these galaxies are detected in the NUV, so either they are intrinsically bright at these wavelengths or, indeed, the highly obscured region is very compact and the UV photons can leak from the less obscured surrounding regions. 

It is important to remark that the discrepancies between the extinctions derived from the Balmer decrements and the IR/UV are expected to be less important at higher redshifts, because the spectral slits collect a more representative part of the total galaxy  light.

\subsection{The BPT diagram}  
\label{sec_bpt}
  
The   \oiii/\hb  versus \nii/\ha  diagram  (Baldwin, Phillips  \& Terlevich~1981; BPT hereafter) provides a useful diagnostics to assess the diversity of the nature of a galaxy population. For example, one can disentangle the existence of nuclear activity from the high \nii/\ha ratios, or select candidates to star-forming/AGN composite systems (e.g. Kauffmann et al.~2003a).

\begin{figure}
\epsscale{0.90}
\plotone{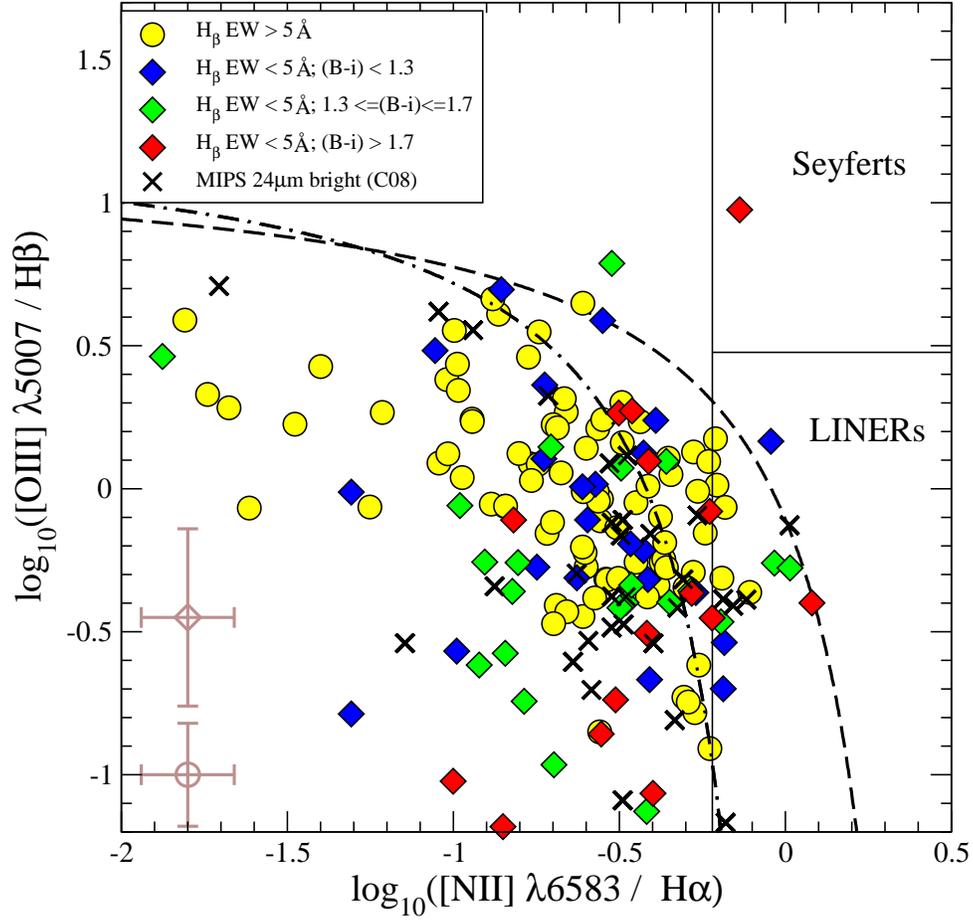}
\caption[]{\label{fig_bpt} The BPT diagram for our \tfm sources at $0.2<z<0.3$. Symbols are the same as in figure \ref{fig_extinct}.  The dashed line indicates the starburst/AGN division line proposed by Kewley et al.~(2001), while the dot-dashed line corresponds to the limit for composite systems (Kauffmann et al.~2003a).}
\end{figure}

 Figure \ref{fig_bpt} shows the BPT diagram for our \tfm galaxies at $0.2<z<0.3$. Symbols are the same as in Figure \ref{fig_extinct}: circles indicate the line ratios of galaxies with both \ha and \hb EW $>5 \rm \, \AA$, while diamonds correspond to galaxies with \ha (\hb) EW $> \, (<) \, 5 \,\rm \AA$ and different $(B-i)$ colours.

  In all cases, the \ha and \hb values are corrected for stellar absorption. However, we do not include extinction corrections. Given the dispersion of these sources in the dust attenuation diagram presented in Figure \ref{fig_extinct}, we preferred to present the crude line ratios in the BPT diagram.  The errors of this approach are basically negligible ($\lsim 0.0025$ and $\lsim 0.02$ on $\log_{10}$ of \nii/\ha and \oiii/\hb, respectively).

 Inspection of Figure \ref{fig_bpt}  shows that \tfm sources occupy quite different places within the BPT diagram, providing evidence of the heterogeneous character of the IR normal galaxy population. We obtained a similar conclusion in C08 for brighter IR sources. 
 
 In our present faint IR sample, we find a significant fraction of galaxies lying in the composite region of the BPT diagram.  This fact suggests that simultaneous star formation and nuclear activity could be present in these sources. These galaxies make around 22\%  of our sources with \ha EW $> 5 \rm \, \AA$ at $0.2<z<0.3$, including $\sim$ 24\% of the sample with \hb EW $>5 \rm \, \AA$, and 20\%  of the sources with \hb EW $< 5 \rm \, \AA$.   In addition, one of the red galaxies with \hb EW $< 5 \rm \, \AA$ is identified as a Seyfert galaxy. 
 
 Within the brighter IR sample (crosses in Figure \ref{fig_bpt}), the percentage of composite sources is $<15$\% (and quite smaller among the sources with \hb EW $> 5 \,\rm \AA$ analysed in C08), and no Seyfert galaxy has been found. Among all the optically-selected zCOSMOS-10k galaxies at similar redshifts, the percentage of composite galaxies in the BPT diagram is only $\sim$10-12\%.  
 
 A bit more than a half of the composite galaxies in our faint IR sample are optically blue ($(B-i)<1.3$), while the rest have redder colours. We note that, in spite of being spectroscopically classified as composites, none of these galaxies has an IRAC power-law SED or is detected in X-rays. This fact suggests that the plausible AGN component is not dominant, and that aperture effects could also have a role in the fraction of sources identified as composite.  
 
 Indeed, if the AGN component were concentrated towards the centre, the line ratios observed in the 1 arcsec-wide optical spectra would not be the same as those obtained from the total galaxy light. In this case, one should probably expect to identify a lower fraction of composite systems among similar IR sources at higher redshifts, for which the central light would not be spectroscopically resolved.

\section{Spectral diagnostics at $\lowercase{z}>0.5$}
\label{sec_linehighz}

\subsection{The mass-metallicity relation}  
\label{sec_metm}

At $z>0.5$, the \ha and \nii emission lines are out of the coverage of the zCOSMOS-bright {\em VIMOS} spectra, but the \oii line starts to be visible instead. In particular, between $z=0.5$ and $z=0.7$, all the three \oii, \hb and \oiii emission lines are located in good regions of the zCOSMOS spectra (i.e. outside the regions severely affected by fringing). The flux line ratios of these three lines can be used to estimate the galaxy oxygen abundances  (with e.g. the so-called $R_{23}$ parameter, i.e. $R_{23}=(\rm [OII] \, \lambda 3727 \,+ [OIII] \, \lambda 4959, 5007)/H\beta$; see Pagel et al.~1979). 

The relation between galaxy mass and metallicity at different redshifts has been widely investigated from the observational (e.g. Lequeux et al.~1979; Zaritsky et al.~1994; Tremonti et al.~2004; Maiolino et al.~2008) and theoretical (e.g. de Lucia et al.~2004; Tissera et al.~2005; de Rossi et al.~2007; Dav\'e \& Oppenheimer~2007; Finlator \& Dav\'e~2008) points of view. The basis of this relation is that galaxy growth and chemical enrichment are closely associated. However, during an episode of star formation, such as those characterising most IR galaxies, gas inflows and outflows can alter the composition of the inter-stellar medium. This fact contributes to the scatter observed in the mass-metallicity relation and, in particular, to the existence of outliers.

In C08, we found that around $1/3$ of the LIRGs and ULIRGs for which we had metallicity measurements at $0.5<z<0.7$ were securely below the mass-metallicity relation. Our aim here is to investigate the fraction of outliers in the mass-metallicity relation among  less luminous IR sources at similar redshifts. Our present faint \tfm sample at $0.5<z<0.7$  spans around a decade in IR luminosities $3 \times 10^{10} \lsim L_{\rm TIR} \lsim 3 \times 10^{11} \, \rm L_\odot$, i.e. it is mainly composed of IR normal galaxies and the less luminous objects among the LIRGs.

We derived metallicities for our 301 IR galaxies with \hb EW $> 5 \,\rm \AA$ and positive \oii and \oiii fluxes. We used  the algorithm described by Maier et al.~(2005), which is based on the Kewley \& Dopita~(2002) model. We applied this algorithm on completely corrected emission lines, i.e. corrected for stellar absorption, aperture effects and extinction. zCOSMOS BLAGN have been excluded from the analysis.

To correct our lines for extinction, we assumed that the sum of the star formation rates (SFRs) derived from the UV and IR  should be equal to that derived from the \hb luminosity, as obtained applying formulae commonly used in the literature (Kennicutt~1998; see C08 for more details).  This procedure could be dubious for some galaxies at low $z\lsim 0.3$ redshifts: we remind that we have concluded in Section \S\ref{sec_dust} that the Balmer decrements and IR/UV attenuations of up to 16\%  of our galaxies at $0.2<z<0.3$ could not be reconciled with any of the commonly used reddening laws.

However, the galaxies in our present sample are at somewhat higher redshifts ($0.5<z<0.7$), where the aperture effects should be less important.  Based on this, and the fact that the typical reddening laws do roughly predict the extinctions of most IR galaxies even at low $z$, we decided to apply the SFR balance procedure --SFR(UV+IR)$\approx$SFR($\rm H_\beta$)-- to estimate the extinction of our IR galaxies at $0.5<z<0.7$. In any case, as we discuss later, this assumption does not affect any of our conclusions.

\begin{figure}
\epsscale{0.60}
\plotone{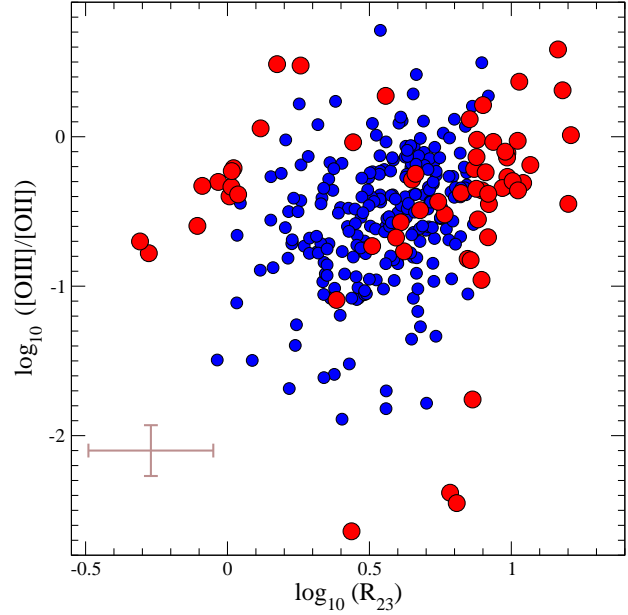}
\caption[]{\label{fig_oiiioiir23} The [OIII]/[OII] ratio versus the $R_{23}$ parameter for the galaxies in our metallicity sample at $0.5<z<0.7$. Small and large circles refer to galaxies that have degenerate and non-degenerate metallicity estimates, respectively.}
\end{figure}

\begin{figure}
\epsscale{0.70}
\plotone{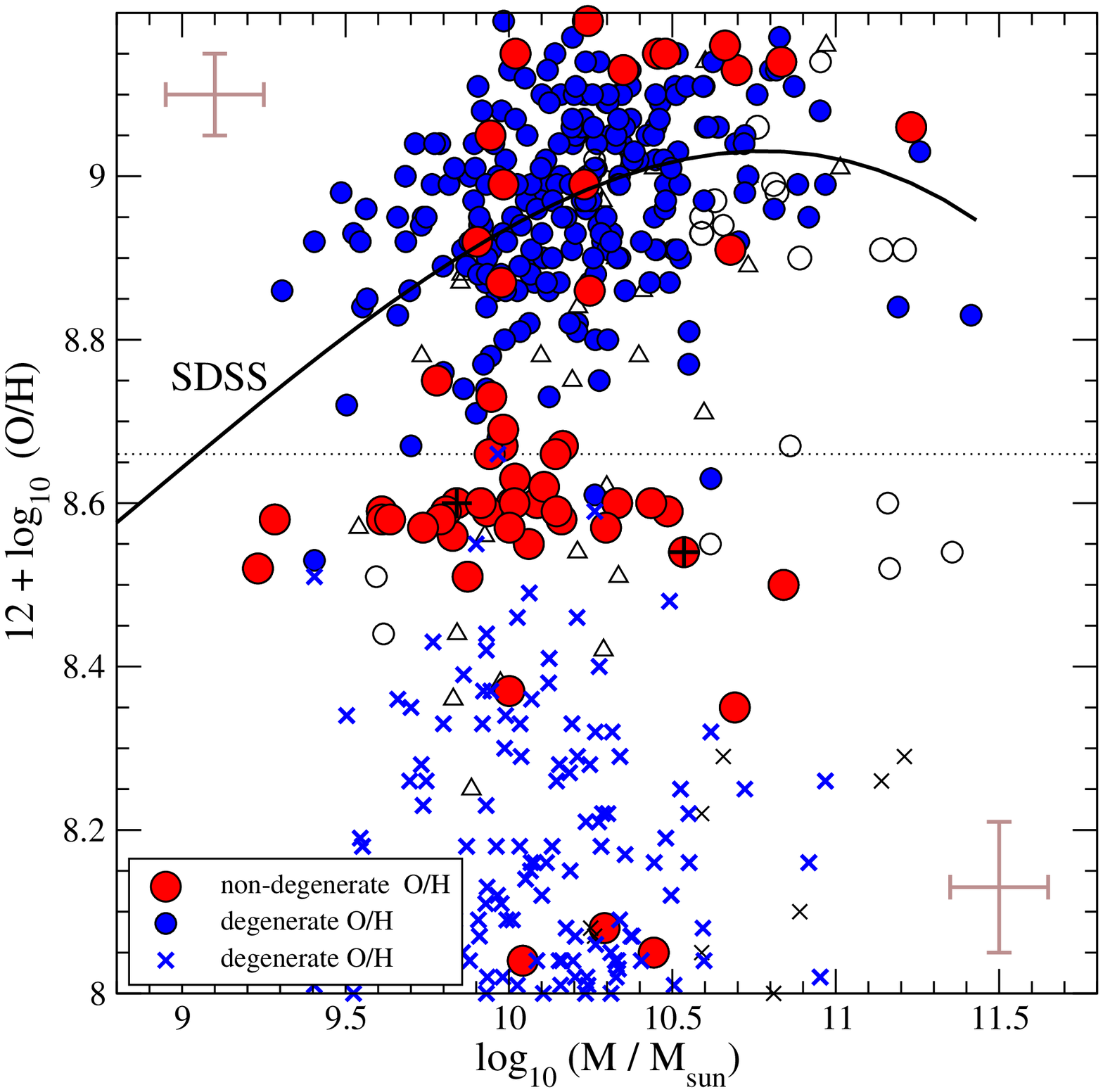}
\caption[]{\label{fig_massmet} Oxygen abundances versus assembled stellar masses for our \tfm galaxies with \hb EW $> 5 \, \rm \AA$ at $0.5<z<0.7$ (filled circles: large and small for non-degenerate and degenerate O/H values, respectively). The thick blue crosses indicate the lower-metallicity alternatives of the blue circles. The plus-like symbols inside the circles refer to X-ray-detected AGN. The average error bars shown in opposite corners of the plot correspond, respectively, to metallicities in the upper and lower branches (non-degenerate metallicities around $12+\log_{10}\rm(O/H) \sim 8.6$ have similar errors to the lower-branch metallicities).  Open circles and thin black crosses correspond to the more luminous \tfm galaxies analysed in C08, while open triangles indicate CFRS galaxies with known metallicities (Maier et al.~2005). The solid line shows the best-fit mass-metallicity relation obtained by Kewley \& Ellison ~(2008) on SDSS galaxies at $z<0.1$. The solar metallicity derived by Asplund et al.~(2004) is also shown for reference (dotted line).}
\end{figure}

Figure \ref{fig_oiiioiir23}  shows the [OIII]/[OII] ratios versus the $R_{23}$-parameter values for the 301 IR galaxies in our metallicity sample. The resulting  metallicity versus stellar-mass diagram is shown in Figure \ref{fig_massmet}.  Stellar masses for all our galaxies have been derived through the synthetic template fitting of their broad-band SEDs, from UV through  near-IR wavelengths (Bolzonella et al.~2009), and correspond to a Salpeter~(1955) initial mass funtion over stellar masses $M=(0.1-100) \, \rm M_\odot$. 
 
 Around 79\% of galaxies in our sample have degenerate $\rm O/H$ values (small blue circles and crosses in Figure \ref{fig_massmet}, showing the upper and lower alternative metallicities, respectively). The metallicity is in general two-valued with respect to the oxygen abundances \oii/\hb and \oiii/\hb, because low line ratios can be produced either by low abundances, or high abundances with very efficient line cooling. 
 
 The remaining $\sim$21\% of our sample have non-degenerate metallicity solutions (large red circles in Figure \ref{fig_massmet}).  Not surprisingly, these galaxies are those with the largest and smallest $R_{23}$ values (see Figure \ref{fig_oiiioiir23}), as the oxygen-abundance versus metallicity curve is single-valued for extreme  $R_{23}$ (this is particularly true for the Kewley \& Dopita calibration; see Kewley \& Dopita~2002).

 From the analysis of Figure \ref{fig_massmet}, we can extract two main conclusions. Firstly, the upper-branch metallicities of our sources with degenerate solutions --whose median redshift is $z\approx0.62$-- concentrate  around the local mass-metallicity relation. As a reference for the local mass-metallicity relation, we considered the results obtained on Sloan Digital Sky Survey galaxies at $z<0.1$  (Kewley \& Ellison~2008; solid line in Figure \ref{fig_massmet}).  The metallicities in this local relation also correspond to  the Kewley \& Dopita~(2002) calibration.  We note that we could not have concluded on this effect in C08,  due to the small number of sources with metallicity measurements in this previous work.

  Secondly,  more than a half of our sources with non-degenerate metallicity solutions lie well below the local mass-metallicity relation, around which the upper-branch solution of our sources with degenerate metallicities are located. Most of the non-degenerate metallicities concentrate around the turning point of the oxygen-abundance versus metallicity $R_{23}$ curve, i.e. at $12+\log_{10}\rm (O/H)\approx 8.6$, consistently with them having large $R_{23}$ values.

  The sources with low non-degenerate metallicities display a variety of stellar masses, putting a secure lower limit of $\sim$ 12\% for the outliers to the mass-metallicity relation within our sample at $0.5<z<0.7$. This percentage is considerably smaller than that estimated for more luminous IR galaxies at similar redshifts (C08), but we emphasize that it is only a lower limit, as there could be more low-metallicity sources among those with degenerate abundances. The under-abundance is probably not the consequence of a line ratio produced by the presence of an AGN within our sources, as only two of these are detected in X-rays (plus-like symbols in Figure \ref{fig_massmet}). 
  
  We observe that a significant outlier ratio to the mass-metallicity relation  has also been reported in optically-selected samples at $z<1$, as e.g. the Canada France Redshift Survey (Lilly et al.~1995) galaxies analysed by Maier et al.~(2005; triangles in Figure \ref{fig_massmet}; see also Lilly et al.~2003). The stellar masses of Maier et al. sources are comparable to those of our \tfm galaxies, suggesting that the   $\sim$ 12\% outlier fraction we derived here is very likely a lower limit.

  To test whether the main conclusions obtained in this Section depend on the adopted dust-extinction-correction recipe, we derived new metallicity values for our galaxies based on the emission lines un-corrected for dust extinction, and compared with our previous results. We found that the percentage of sources with non-degenerate metallicity values in the un-corrected case is 19\%, and the upper branch solutions in the degenerate cases are still compatible with the local mass-metallicity relation within the error bars. The percentage of secure outliers to this relation is $\sim 9$\%.  These figures are very similar to those obtained using the extinction-corrected emission lines.

\subsection{The IR phase in the galaxy star-formation history}
\label{sec_irph}

\subsubsection{Individual $D_n(4000)$ measurements}
\label{sec_irphind}

The $4000 \, \rm \AA$-break in galaxy spectra is produced by the differential contribution of  young and old stars that, respectively, dominate the spectral light bluewards and redwards of $\lambda= 4000 \, \rm \AA$. The strength of this feature $\rm D_n(4000)$, usually defined as the ratio of the average flux densities between the $(4000-4100)$ and $(3850-3950) \, \rm \AA$ bands (Balogh et al.~1999), is a very good indicator of the time passed since the last significant burst of star formation occurred in the galaxy.

Besides, the presence of  higher-order Balmer lines in absorption, such as $\rm H_\delta \, \lambda 4102$, indicates recent star formation activity, which typically has occurred within the last Gyr (e.g. Dressler \& Gunn 1983). The combined diagnostics of the $\rm D_n(4000)$ parameter and $\rm H_\delta$ EW in optical spectra is, then, commonly used as a powerful tool to constrain the galaxy star formation history (see e.g. Kauffmann et al.~2003b; C08).

Our results in C08 showed that the vast majority of optically bright ULIRGs and some of the most luminous LIRGs at $0.6<z<1.0$ were characterised by spectra with $\rm D_n(4000) \lsim 1.2$. We interpreted this effect as the consequence of secondary bursts of star formation produced in galaxies with underlying old stellar populations. Here, we do a combined analysis of the IR-brighter C08 sample and our current fainter sample to investigate whether this mode of star formation is similarly valid for galaxies of different IR luminosities.

Our present \tfm sample at $0.6<z<1.0$ is mainly composed of galaxies with rest-frame \tfm luminosities $5 \times 10^9 \lsim \nu L_\nu^{\rm 24 \, \mu m} \lsim  10^{11} \, \rm L_\odot$, i.e. total IR luminosities $5.3 \times 10^{10} \lsim L_{\rm TIR} \lsim 5 \times 10^{11} \, \rm L_\odot$. This range is particularly important for IR galaxies at $0.6<z<1.0$, as it contains most of the IR luminosity density at these redshifts  (the total IR luminosity function is characterised by $L^\ast \approx 2.5 \times 10^{11} \, \rm L_\odot$ at $z=1$; see Caputi et al.~2007). The brighter C08 sample extends the IR luminosity range to $\nu L_\nu^{\rm 24 \, \mu m} \approx 5 \times 10^{11} \, \rm L_\odot$, i.e. $L_{\rm TIR} \approx 1.5 \times 10^{12} \, \rm L_\odot$. Thus, the combined analysis we perform here covers two decades in mid-IR luminosities (or 1.5 decades in $L_{\rm TIR}$).

\begin{figure}
\epsscale{0.80}
\plotone{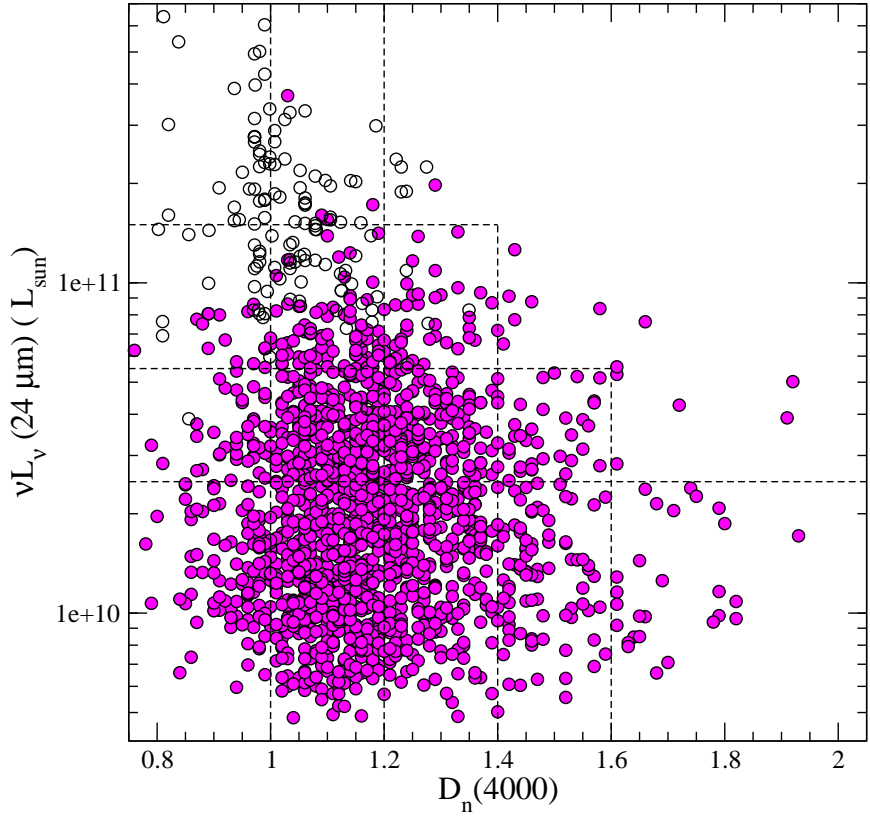}
\caption[]{\label{fig_l24dn4000} Rest-frame \tfm luminosities versus $\rm D_n(4000)$ parameter values for our \tfm galaxies at $0.6<z<1.0$. Filled circles correspond to galaxies in our current faint \tfm sample, while open circles indicate the C08 \tfm sources. The dashed lines delimit the bins we have adopted for spectral stacking and  $\rm H\delta$ EW versus $\rm D_n(4000)$ measurements (cf. Figure \ref{fig_hddn4}). The $\rm D_n(4000)$ values in this figure are not corrected for extinction.}
\end{figure}

We measured the $\rm D_n(4000)$ parameter values on the spectra of a total of 1722 IR sources at $0.6<z<1.0$. This figure excludes BLAGN. The resulting $\nu L_\nu^{\rm 24 \, \mu m}$ versus $\rm D_n(4000)$ diagram is shown in Figure \ref{fig_l24dn4000}. In this figure we can see that, in contrast to the more luminous IR galaxies studied in C08, the less luminous sources display a wide range of $\rm D_n(4000)$ values, from $\approx 0.8$ to $\approx 1.8$. Still, a bit more than 60\% of them have spectra with $\rm D_n(4000)<1.2$. 

We used the diagram in Figure \ref{fig_l24dn4000} to classify our galaxies in different  bins of $\rm D_n(4000)$  values and \nLntfm luminosities. We stacked the zCOSMOS spectra of all galaxies in each bin, in order to measure their average spectral properties (cf. Section \S\ref{sec_irphstack}). The  bins have been chosen in such a way to produce a good sampling of the ($\rm D_n(4000)$; \nLntfm) space with a sufficient number of objects.

The $\rm D_n(4000)$ values shown in Figure \ref{fig_l24dn4000} and used for the classification are not corrected for dust extinction.  This assures that we  stacked spectra with similar characteristics prior to any dust extinction assumption. In any case, the impact of the dust extinction corrections is small: for example, an extinction $A_V=2$  only produces a correction of $\sim$8\% on the $\rm D_n(4000)$ values.

\subsubsection{The $H_\delta$ EW - $D_n(4000)$ diagram: discerning the modes of star formation}
\label{sec_irphstack}

 Locating our galaxies in the $\rm H_\delta \, \lambda 4102 \,$ EW - $\rm D_n(4000)$  diagram allows us to have a better insight of the time at which the IR phase occurs within the galaxy star formation history.

As we explained in C08, the $\rm H_\delta$ absorption EWs in the zCOSMOS spectra of IR galaxies are usually quite small and the associated errors too large as to rely on the individual measurements. Therefore, to perform  $\rm H_\delta$ EW measurements with more precision, we stacked the IR galaxy spectra.

 We classified all (bright and faint) \tfm galaxies at $0.6<z<1.0$ in 14 groups according to their  $\rm D_n(4000)$ and \nLntfm values, and averaged their respective spectra.  We then measured the  $\rm D_n(4000)$  parameter and $\rm H_\delta$ EW on each composite spectrum. The number of galaxies in each stack varies between 495 and 28.  We have not separated our galaxies according to their redshifts because the binning in the $0.6<z<1.0$ redshift range does not produce any relevant effect in all the analysis presented here.

\begin{figure}
\epsscale{1.0}
\plotone{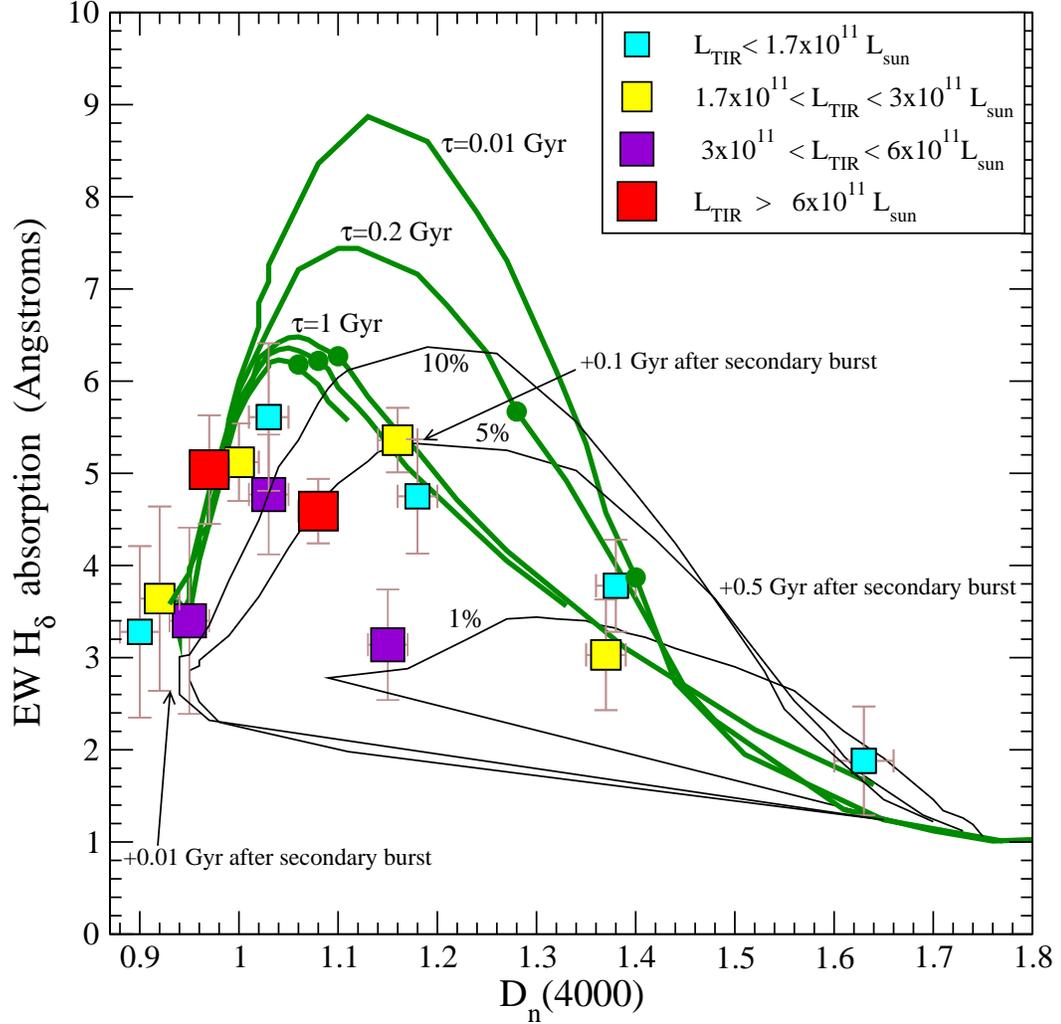}
\caption[]{\label{fig_hddn4} The $\rm H\delta$ EW versus $\rm D_n(4000)$ diagram. The average location of our zCOSMOS \tfm galaxies at $0.6<z<1.0$, classified in intervals of mid-IR luminosity \nLntfm and  $\rm D_n(4000)$ values, are indicated by filled squares (note: the average $\rm D_n(4000)$ values shown in this plot are statistically corrected for dust extinction; see text).  The total IR luminosities $L_{\rm TIR}$ shown in the labels have been obtained applying the conversion formula $L_{\rm TIR}=6856 \times (\nu L_\nu^{\rm 24 \, \mu m})^{0.71}$ (Bavouzet et al.~2008). The thick lines show the tracks produced by various exponentially-declining star formation histories: from top to bottom, $\tau=0.01$, 0.2, 1, $2 \, \rm Gyr$ and $\tau \rightarrow \infty$ (i.e. constant SFR).  Tracks only show the evolution of each model up to 6 Gyr, which is approximately the age of the Universe at $z=1$. The filled circle intersecting each track marks the position at 1 Gyr. Thin solid lines indicate the effect of secondary bursts of different amplitudes (the relative amplitude with respect to the primary burst is indicated with a percentage), produced at different ages of an underlying old stellar population.}
\end{figure}

 The resulting $\rm H_\delta$ EW - $\rm D_n(4000)$  diagram for our stacked spectra is shown in Figure \ref{fig_hddn4}. Positive EW values in this diagram indicate $\rm H_\delta$ in absorption, and are corrected for line filling. To do this correction, we  measured the $\rm H\gamma$ emission line  on each composite, corrected it for stellar absorption, and assumed an intrinsic decrement $\rm H\gamma/H\delta=1.80$ (for a case B recombination with temperature $T=10,000 \rm K$; Osterbrock~1989).  The final $\rm H_\delta$ EW values carry significant error bars, mainly because the crudely measured values are quite small and also because of the uncertainties in the corrections. The $\rm D_n(4000)$ values shown in Figure \ref{fig_hddn4} do include dust extinction corrections, which we estimated  using the median extinction of the galaxies in each stack group. The total IR luminosities shown in the labels have been obtained by converting the limits of the \nLntfm bins with $L_{TIR}=6858 \times (\nu L_\nu^{\rm 24 \, \mu m})^{0.71}$ (Bavouzet et al.~2008). One should keep in mind that using a different recipe to convert mid-IR into total IR luminosities will produce some variations in the derived $L_{TIR}$ values.

 The tracks shown with thick solid lines in Figure \ref{fig_hddn4} correspond to galaxy models with different exponentially-declining star formation rates SFR $\propto e^{(-t/\tau)}$, where $\tau$ is the characteristic decaying time that  varies between $\tau=0.01 \, \rm Gyr$ and $\tau \rightarrow \infty$   (i.e. a model with a constant SFR).  We have produced these tracks by measuring the  $\rm H_\delta$ EW and $\rm D_n(4000)$ parameter on galaxy synthetic SEDs generated with the GALAXEV code and the 2007 version of the Bruzual \& Charlot models (Bruzual \& Charlot~2003; Bruzual~2007). All these tracks are plotted only up to ages $\leq$ 6 Gyr, as this is approximately the age of the Universe at $z=1$.

 As we have discussed in C08, the average position of the most luminous IR galaxies in the $\rm H_\delta$ EW - $\rm D_n(4000)$  diagram cannot  be reproduced by any of these simple $\tau$-decaying star formation histories. Our present analysis confirms that this is the case for galaxies with total IR luminosities $L_{\rm TIR} > 3\times 10^{11} \, \rm L_\odot$ (squares with the two largest sizes in Figure \ref{fig_hddn4}). Instead, secondary bursts of star formation in galaxies with underlying old ($\gsim 3 \, \rm Gyr$) stellar populations appear as a plausible mechanism to explain the spectral characteristics of these most luminous IR sources (thin solid lines in Figure \ref{fig_hddn4}). We note that bursts producing 5-10\% of the galaxy stellar mass are sufficient to drastically move the position of a few-Gyr-old galaxy in the diagram from $\rm D_n(4000) \gsim 1.5$ to $\rm D_n(4000)<1.2$.

 Interestingly, although the average ($\rm H_\delta$ EW;$\rm D_n(4000)$)  points for the most luminous IR galaxies with $\rm D_n(4000)<1$ do seem to be consistent with  $\tau$-decaying  models, they intersect their tracks at ages $<0.1$ Gyr. As it is quite unlikely that there exist many of such young galaxies at $0.6<z<1.0$, the scenario of secondary bursts of star formation is still the most reasonable to explain the properties of these sources. This argument is supported by the fact that the stacked spectra of these galaxies show some signatures of old  underlying stellar populations, such as the presence of CaII H \& K absorption lines. Also, these sources are known to have already assembled intermediate-to-large stellar masses ($M \gsim 1 \times 10^{10} \, \rm M_{\odot}$; see e.g. Caputi et al.~2006a), suggesting that we are not witnessing their first episode of star formation.

 As we discussed before, for the less luminous IR ($L_{\rm TIR} < 3\times 10^{11} \, \rm L_\odot$) galaxies (squares of the two smallest sizes in Figure \ref{fig_hddn4}), the $\rm D_n(4000)$ parameter has a wider range of values.  Taking into account their average position in the $\rm H_\delta$ EW - $\rm D_n(4000)$ diagram and other features in the stacked spectra, we can obtain clues on the star formation histories of the sources in different $\rm D_n(4000)$ bins.

\begin{figure}
\epsscale{1.0}
\plotone{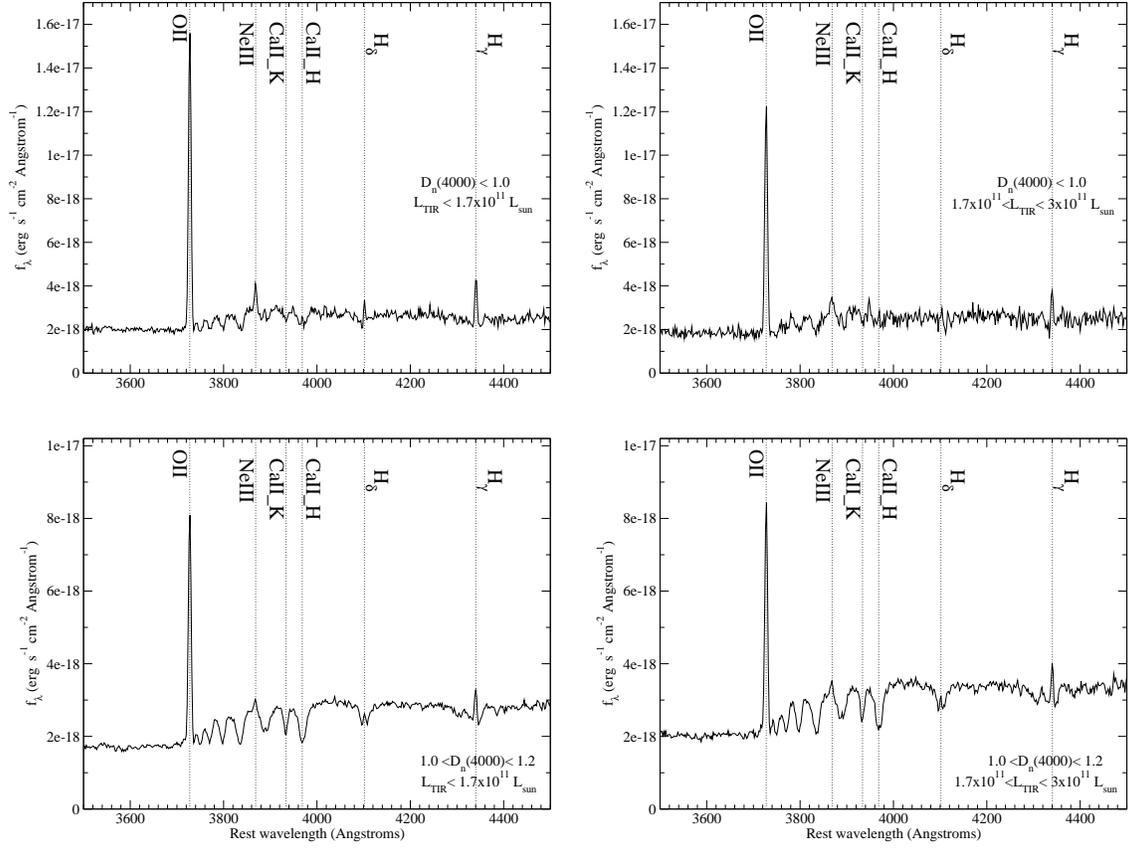}
\caption[]{\label{fig_stdn4lt12} Stacked spectra for the less luminous IR galaxies in our sample, and the smallest $\rm D_n(4000)$ values (the bins displayed in the labels are those used for the stacks, i.e. $\rm D_n(4000)$ is not corrected for extinction).}
\end{figure}

 The sources with $\rm D_n(4000)<1$ are characterised by an average spectra with relatively weak absorption lines, and an $\rm H_\delta$ line with EW $< 2\, \rm \AA$ (that, in the case of the less luminous IR galaxies, is actually   observed in emission; see upper panels in Figure \ref{fig_stdn4lt12}). They also show a prominent [OII] emission line and, those with $L_{\rm TIR}<1.7\times 10^{11} \, \rm L_\odot$, a modest but significant [NeIII] line. These spectral characteristics, and the resulting average position of these sources in the $\rm H_\delta$ EW -$\rm D_n(4000)$ diagram, suggest that these are either really young galaxies or, more likely, they are experiencing a major secondary burst of star formation that hides any plausible older stellar component. Note that the ionisation lines could also be the consequence of the presence of AGN. However, only one of these sources is detected in the X-ray catalogues for the COSMOS field, and only two have an IRAC power-law SED (one of which is the X-ray source).

 The sources with $L_{\rm TIR} < 3\times 10^{11} \, \rm L_\odot$ in the $1.0<\rm D_n(4000)<1.2$ bin are characterised by quite different average stacked spectra (lower panels in Figure \ref{fig_stdn4lt12}). These galaxies display prominent CaII H \& K absorption lines, indicating that the underlying old stellar component is important. Their average position in the diagram at (extinction-corrected) $\rm D_n(4000) \approx 1.02$  seems compatible with some of the $\tau$-decaying  models at young ($<1$ Gyr) ages. However, the optical spectra indicate that these galaxies are very probably older.  Thus, once more, secondary bursts are the most plausible mechanism for the on-going star formation in these objects.

\begin{figure}
\epsscale{0.60}
\plotone{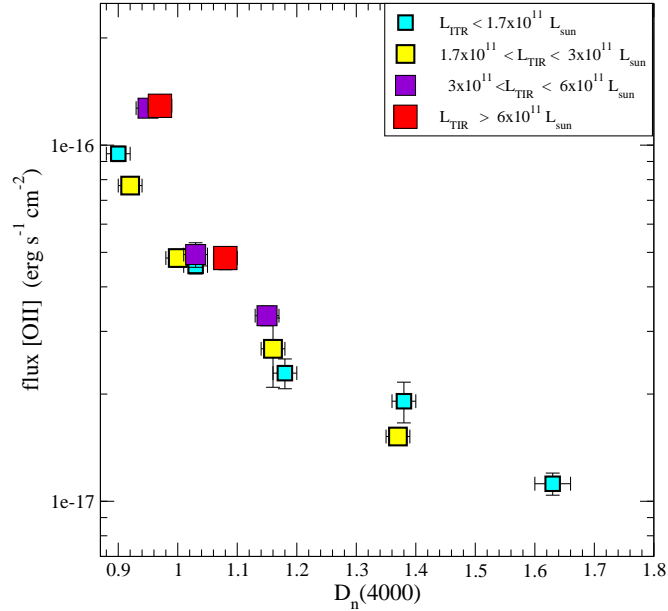}
\caption[]{\label{fig_oiidn4} The average [OII] fluxes versus $\rm D_n(4000)$ values measured on our stacked spectra at $0.6<z<1.0$, corresponding to different \nLntfm and $\rm D_n(4000)$ bins. Symbols are the same as in Figure \ref{fig_hddn4}.}
\end{figure}

 The stacking points indicating $L_{\rm TIR} < 3\times 10^{11} \, \rm L_\odot$ galaxies at $\rm D_n(4000) > 1.15$ in Figure \ref{fig_hddn4} correspond to a different situation. All these points intersect the $\tau$-decaying  model tracks at ages $>1$ Gyr, so such models allow for the underlying old stellar component that is present in these galaxies. At the same time, although the SFR is exponentially declining,  it is still sufficient to explain the observed average [OII] fluxes (cf. Figure \ref{fig_oiidn4}). For example, the stacking point for $L_{\rm TIR} < 1.7\times 10^{11} \, \rm L_\odot$ galaxies at  $\rm D_n(4000)=1.38$ and $\rm H_\delta$ EW$=3.78 \, \rm \AA$ is consistent with a $\tau=0.2$ Gyr model at an age  $\sim 1.4$ Gyr. By that time, the initial SFR has decayed by a factor of $e^{-7} \approx 10^{-3}$. This situation is feasible: a galaxy that started its life being a ULIRG would still be producing a few $\rm M_\odot/yr$ 1.4 Gyr later.
 
 Also, the stacking point at $\rm D_n(4000)=1.63$ could be well explained by the model of a 6 Gyr-old galaxy with a much slower SFR decaying time (see the $\tau=1$ Gyr track in Figure \ref{fig_hddn4}). In spite of being quite old, the SFR of such a galaxy is still non-negligible  ($e^{-6} \approx 2.5 \times 10^{-3}$  times its initial value). Thus, this model simultaneously accounts for the absorption lines and [OII] emission observed in the average spectrum.
 
 It is important to remark that the fact that $\tau$-decaying models are able to explain the spectral characteristics of these less luminous IR galaxies, the secondary-burst scenario cannot be excluded. In all cases, one can find the right combination of  burst amplitude and onset age (i.e. the age of the galaxy when the burst is produced), in order to reproduce the average location of galaxies with different  
 $\rm D_n(4000)$ values in the $\rm H_\delta$ EW - $\rm D_n(4000)$ diagram.

 The IR galaxies for which $\tau$-decaying models offer a good explanation for the on-going star formation (i.e. those with un-corrected $\rm D_n(4000) >1.2$) constitute $\sim$ 40\% of our $L_{\rm TIR} < 3\times 10^{11} \, \rm L_\odot$ sample at $0.6<z<1.0$. This result clearly shows that a quiescent mode of star formation can drive the IR emission in a significant fraction of the less luminous LIRGs and IR normal galaxies.

\subsection{Candidates to young starbursts}
\label{sec_young}

Although the average spectra of most \tfm galaxies at $0.6<z<1.0$ are characterised by having high-order Balmer lines such as $\rm H_\delta$ in absorption, in some objects those lines are present in emission.   A significant $\rm H_\delta$ emission is expected for galaxies at the very early stages ($\sim 10^7$ years) of a new burst of star formation, when the  $\rm H_\delta$ absorption associated with intermediate-age stars is still not sufficient to compensate the emission  produced by the youngest ones.

As we explained in C08 and in Section \ref{sec_irphstack} of this present paper, the  $\rm H_\delta$ EW measurements on the spectra of our IR galaxies generally have too low signal-to-noise ratios as to safely consider them on an individual basis.  In spite of this fact, we used our tentative $\rm H_\delta$ EW measurements and secure $\rm D_n(4000)$ values to select galaxies that are candidates to be hosting young starbursts.

Our present sample of 1587 faint \tfm galaxies at  $0.6<z<1.0$ contains 31 galaxies with $\rm D_n(4000)<1.0$ that have $\rm H_\delta$ EW$> 5\, \rm \AA$ and $0 < d(\rm H_\delta)/ \rm H_\delta < 1$ (i.e. $\rm H_\delta$ is securely in emission). These galaxies constitute 18\% of all our faint sources with $\rm D_n(4000)<1.0$ at  $0.6<z<1.0$, and around 3\% of those that are within $\sim 10^8$ yr of a new burst of star formation (cf. Figure \ref{fig_hddn4}). So, the clear cases of galaxies hosting very young starbursts constitute a  minor fraction of our IR sample (we extracted a similar conclusion for brighter IR sources in C08).

There are two alternative explanations for this situation. One possibility is that the IR phase is not equally likely within the first $\sim 10^8$ yr of a new burst, i.e.  galaxies containing very young bursts are still too faint to be detected at IR wavelengths.
The other possibility is that the earliest stages of star formation are heavily obscured. If only the gas had a very  large extinction, the observed $\rm H_\delta$ emission would be weak, and the $\rm H_\delta$ EW$> 5\, \rm \AA$ cut could be missing young starburts. If both the gas and stars were very obscured, the galaxies would look faint in the optical wavebands and, in that case, they could simply be missed by the zCOSMOS-bright  $I<22.5$ AB mag selection.

\section{Summary and Conclusions}
\label{sec_concl}

 In this paper, we have identified 3244 \tfm-selected galaxies with $0.06< S_{24 \, \rm \mu m}\lsim 0.50 \, \rm mJy$  within the zCOSMOS-bright 10k sample, and studied different optical spectral properties depending on their redshifts. Our sample spans e.g. total IR luminosities $\sim 3 \times 10^{10}< L_{\rm TIR} < 9 \times 10^{10} \, \rm L_\odot$ at $z=0.5$, and  $\sim9 \times 10^{10} < L_{\rm TIR} < 5 \times 10^{11} \, \rm L_\odot$ at $z=1$. This means that our sources are IR normal galaxies at $z<0.5$, and a mixture of IR normal galaxies and LIRGs at $0.5<z<1.0$. This works complements our previous study of the zCOSMOS-bright spectra of more luminous IR galaxies at similar redshifts (C08). The fainter luminosity range studied here is particularly important, as it contains the bulk of the IR background energy at these redshifts.
 
 For galaxies at $0.2<z<0.3$, we took advantage of the simultaneous presence of the \ha and \hb lines in the zCOSMOS spectra to study the relation between line emission, broad-band colours and dust extinction.   We also discussed the possibility for the IR emission to be associated with dust heated by old rather than young stellar populations.  We summarise our results as follows:
 
$\bullet$   92\% of our \tfm galaxies at $0.2<z<0.3$ have \ha EW$> 5 \, \rm \AA$, which indicates  that the vast majority of IR sources at these redshifts are associated with significant on-going star formation. 

In most of these sources, the relation between the gas obscuration, i.e. $L(\rm H\alpha)/L(\rm H\beta)$, and IR/UV attenuation can be reasonably explained with different reddening laws of common use in the literature. However, for up to 16\% of our sample at $0.2<z<0.3$, the Balmer decrement is much larger than predicted by any of these laws. We believe that this is produced by a combination of an inhomogenous dust distribution and the fact that the zCOSMOS spectra only receive the light from a 1-arcsec width slit centred at the source.  The presence of a very obscured star-forming region towards the centre can produce very high $L(\rm H\alpha)/L(\rm H\beta)$ ratios.   We note that this effect should be much less important at higher redshifts, where the 1-arcsec width covered by the {\em VIMOS} slit collects a more representative fraction of the total  galaxy light. 
 
It is necessary to emphasize that, although the  broad-band SEDs of many low-redshift IR galaxies are dominated by old stellar populations, the optical spectra reveal the presence of on-going star formation in most cases. Thus, for these galaxies, one cannot conclude that the IR light is produced by  dust heated by old stars. More likely, the young stellar component is the main energy source for the IR emission.  Of course, this does not exclude the possibility that the old stars have a minor contribution to the dust heating.

$\bullet$ Although there is no one-to-one relation between the \ha and \hb emission and the galaxy optical broad-band colours, around 80\% of the objects with \ha and \hb EW $> 5 \, \rm \AA$  are in the blue ($B-i<1.3$) cloud, and around 70\% of those with \hb EW $< 5 \, \rm \AA$ are in the `green valley' or redder ($B-i \geq 1.3$).

$\bullet$  The location in the BPT diagram of our galaxies with \ha EW $> 5 \, \rm \AA$  suggests that $\sim 22\%$ of them are composite systems. However, in none of these galaxies can the presence of an active nucleus be confirmed with  the X-ray or  {\em IRAC} data available in the COSMOS field. This fact suggests that, although an AGN component might be present in these galaxies, in any case is not dominant, and that aperture effects could also influence the fraction of sources identified as composites.

$\bullet$  We identified only 5 galaxies in our sample at $0.2<z<0.3$ which have no spectral signs of star formation and, besides, have elliptical or lenticular morphologies. We believe that these are the only candidates  in our sample at $0.2<z<0.3$  for which most of the IR emission could genuinely be produced by dust heated by old rather than new stellar populations.

\vspace{0.5cm}  
 
  The second part of our analysis has been devoted to galaxies at $z>0.5$. We used the zCOSMOS spectra to determine oxygen abundances, measure the  $\rm D_n(4000)$ parameter and the average $\rm H_\delta$ EWs. Our results are the following:

$\bullet$  79\% of our galaxies with \hb EW $> 5 \, \rm \AA$ at $0.5<z<0.7$ have degenerate metallicity values. This is a consequence of estimating metallicities based on only three emission lines (through the $R_{23}$ parameter).  When we place these galaxies in a stellar mass-metallicity diagram, we see that the upper-branch solution is compatible with the mass-metallicity relation in the local Universe within the error bars. This implies that our data do not allow us to conclude on any evolution of the mass-metallicity relation between $z\sim0$ and $z\sim0.6$. Among the 21\% of galaxies with non-degenerate metallicity measurements, we find that more than a half are well below the mass-metallicity relation. Over our total IR-galaxy population at $0.5<z<0.7$, they put a secure lower limit of 12\% to the fraction of outliers to this relation. This percentage is quite smaller than that observed for LIRGs and ULIRGs by C08. However, the comparison of our results with those for optically-selected galaxies with similar stellar masses and well-determined metallicities suggests that the 12\% fraction is indeed a lower limit.

$\bullet$   In contrast to the results for more luminous IR galaxies, we find that our present sample at $0.6<z<1.0$ displays a wide range of $\rm D_n(4000)$ values.  Still, $\sim 60 \, (80)$\% of our sources have un-corrected (corrected) $\rm D_n(4000)<1.2$.

 $\bullet$  A combined analysis of our present faint IR sample and previous C08 brighter sample at $0.6<z<1.0$ has allowed us to investigate the possible modes of star formation in IR galaxies. The burst-like mode of star formation is necessary to simultaneously explain the different average spectral properties of the most luminous IR galaxies ($L_{TIR} \gsim 3\times 10^{11} \, \rm L_\odot$). Instead, for nearly 40\% of the less luminous LIRGs and most luminous IR normal galaxies, a continuous, quiescent star-formation mode is possible. Realistically, the $L_{\rm TIR}$ for the transition at $0.6<z<1.0$ should be considered to be $(3\pm2)\times 10^{11} \, \rm L_\odot$, to take into account the errors in the determination of the total IR luminosities and the data binning. 
 
 Our findings are consistent with other recent results in the literature, which indicate that around a half of IR galaxies of intermediate luminosities at $z\sim1$ are typical disks that do not show any sign of disturbance (see e.g. Elbaz et al.~2007). They could also be explained in the scenario proposed by theoretical studies that suggest that some galaxies could have had a star formation history characterised by steady gas flows since early times (Dekel et al.~2009).

  In summary, our optical spectral analysis has clearly shown the diverse nature of the galaxies that make the bulk of the mid-IR background at $z<1$. And, even when star formation seems to be the most important driver of the IR emission at these redshifts,  IR galaxies are characterised by a complex variety of star formation histories.

\acknowledgments

This paper is based on observations made with the {\em VIMOS} spectrograph on the {\em Melipal-VLT} telescope, undertaken at the European Southern Observatory (ESO) under Large Program 175.A-0839. Also based on observations made with the {\em Spitzer} Observatory, which is operated by the Jet Propulsion Laboratory, California Institute of Technology, under NASA contract 1407. 
We would like to thank the anonymous referee for a constructive report of this paper.

%

\clearpage

%

%

%

%

\end{document}